\documentclass[12pt]{iopart}

\usepackage{graphicx}
\usepackage{epsfig}

\usepackage{amssymb}
\usepackage{color}
 \expandafter\let\csname equation*\endcsname\relax
  \expandafter\let\csname endequation*\endcsname\relax
\usepackage{amsthm}
\usepackage{amsmath,bbm}

\usepackage{graphicx}
\usepackage{dcolumn}
\usepackage{bm}
\usepackage{psfrag}
\usepackage{epsfig}
\usepackage{amsmath} 
\usepackage{amssymb}
\usepackage{color}
\usepackage[FIGTOPCAP,raggedright,nooneline]{subfigure}
\newcommand{\appropto}{\mathrel{\vcenter{
  \offinterlineskip\halign{\hfil$##$\cr
    \propto\cr\noalign{\kern2pt}\sim\cr\noalign{\kern-2pt}}}}}
\newcommand{\bra}[1]{\langle #1 |}
\newcommand{\ket}[1]{| #1 \rangle}

\newcommand{\braket}[2]{\ensuremath{\langle#1|#2\rangle}}

\newcommand{\id}{{\sf 1 \hspace{-0.3ex} \rule{0.1ex}{1.52ex}\rule[-.01ex]{0.3ex}{0.1ex}}}
\newcommand{\ignore}[1]{}

\newcommand{\I}{\ensuremath{\mathrm{i}}}
\begin{document}
\title{Exploring unresolved sideband, optomechanical strong coupling using a single atom coupled to a cavity}
\author{Lukas Neumeier$^1$, Darrick E. Chang$^{1,2}$}
\address{$^1$ICFO-Institut de Ciencies Fotoniques, The Barcelona Institute of Science and Technology, 08860 Castelldefels, Barcelona, Spain, $^2$ICREA-Instituci\'{o} Catalana de Recerca i Estudis Avan\c{c}ats, 08015 Barcelona, Spain}
\ead{lukas-neumeier@gmx.de}
\date{\today}

\begin{abstract}
A major trend within the field of cavity QED is to boost the interaction strength between the cavity field and the atomic internal degrees of freedom of the trapped atom by decreasing the mode volume of the cavity. In such systems, it is natural to achieve strong atom-cavity coupling, where the coherent interaction strength exceeds the cavity linewidth, while the linewidth exceeds the atomic trap frequency. While most work focuses on coupling of photons to the internal degrees of freedom, additional rich dynamics can occur by considering the atomic motional degree of freedom as well. In particular, we show that such a system is a natural candidate to explore an interesting regime of quantum optomechanics, where the zero-point atomic motion yields a cavity frequency shift larger than its linewidth (so-called single-photon optomechanical strong coupling), but simultaneously where the motional frequency cannot be resolved by the cavity. We show that this regime can result in a number of remarkable phenomena, such as strong entanglement between the atomic wave-function and the scattering properties of single incident photons, or an anomalous mechanism where the atomic motion can significantly heat up due to single-photon scattering, even if the atom is trapped tightly within the Lamb-Dicke limit.
\end{abstract}
\maketitle
%
%
In optomechanics much progress has been made improving the control over the interaction between photons and phonons at the quantum level \cite{aspel}. Lately there have been many important experimental successes, which include the generation of slow light with optomechanics \cite{transp}, the entanglement of motion with microwave fields \cite{entangling}, and very recently remote entanglement between two micromechanical oscillators \cite{entanglement}.
 For most of the quantum phenomena observed thus far or envisioned, sideband resolution, where the mechanical frequency $\omega_m$ exceeds the cavity linewidth $\kappa$, is required. For example, this enables cooling to the quantum ground state \cite{cooling1,cooling2}, which represents a fiducial pure state preparation. In one remarkable theoretical work \cite{rabl}, it has been predicted that the combination of sideband resolution and single-photon optomechanical strong coupling -- where the zero-point motional uncertainty induces a shift in the optical resonance frequency larger than the cavity linewidth -- would enable the generation of non-classical, anti-bunched light.

Here, we study the complementary regime of single-photon optomechanical strong coupling, but with unresolved sidebands \cite{opto1,opto2}. We show that interesting quantum effects both in the light and motion can be observed, at least when the mechanical system is well-isolated and can be separately prepared in the ground state. A natural candidate system consists of a single atom \cite{photonblockadekimble,rempe,rempe2,reiserer,rauschen,rooting} or ion \cite{ion1,ion2,ion3,ion4,ion5,ion6}
in cavity QED, whose electronic transition is strongly coupled to a near-resonant optical mode. To provide an intuitive picture, strong coupling within cavity QED \cite{strong,strong2} implies that a point-like atom produces a shift in the cavity resonance frequency that is larger than the cavity linewidth, when the atom is situated at a cavity anti-node. If the atom is displaced by a quarter wavelength to a node, this shift vanishes. Given the light mass, it is straightforward for a trapped atom to have a zero-point motion on that scale, thus realizing single-photon optomechanical strong coupling. Furthermore, realistic trap frequencies for atoms are quite low ($\lesssim \mathrm{MHz}$), and are naturally exceeded by the cavity linewidth for small cavities \cite{rauschen,rooting,lukin,day}. In this regime of optomechanical strong coupling and unresolved sidebands, the interesting physics arises because the resonance frequency of the cavity correlates strongly with the atomic position, and as the reflection or transmission of a single photon depends on the resonance frequency, a strong entanglement between photon and motion ensues, which is visible in both of these degrees of freedom.

In this work we begin by considering a single atom externally trapped inside a cavity mode that is driven with a coherent state. 
When the cavity frequency is detuned from the atomic resonance, we derive from the full Jaynes-Cummings model of cavity QED an effective optomechanical Hamiltonian, which only depends on the atomic motion and cavity degrees of freedom.
We proceed by tracing out the cavity degree of freedom and analytically derive an effective quantum master equation describing the motional dynamics of the atom only.
This master equation would allow for the calculation of motional energy eigenvalues and their lifetimes, and yields interesting insights in the heating processes associated with entanglement between light and motion. This entanglement is also directly revealed by applying scattering theory to exactly solve for the joint atom-photon wave function following the scattering of a single incident photon. Using this formalism, we show that the properties of the scattered photon can become entangled with the atomic motion on length scales much smaller than either the resonant wavelength or the atomic zero-point motion. As one consequence, once the photon is traced out, the atomic motion is seen to heat up significantly, even if the atom is tightly trapped within the Lamb-Dicke limit. We also show that this entanglement can manifest itself in the second-order correlation functions of the outgoing field given a weak coherent state input, or be used to produce a heralded single-phonon Fock state of the atomic motion.
\begin{figure}
\centering
\includegraphics[width=0.6\textwidth]{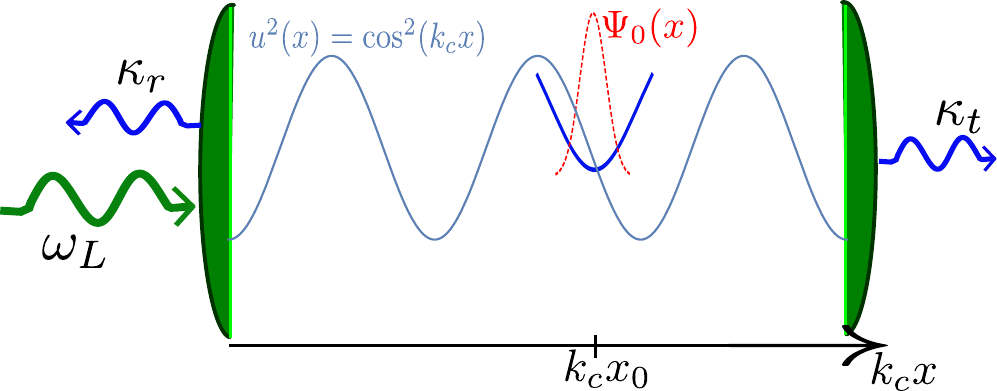} 
\caption{An atom is trapped externally by a potential (blue) with equilibrium position $x_0$ inside a cavity with intensity mode profile $u^2(x)$. $\Psi_0(x)$ is the initial wave function of the atomic motion. Incident photons with frequency $\omega_L$ arrive from the left. The left mirror has a decay rate of $\kappa_r$ and the right mirror has a decay rate of $\kappa_t$.   
}
\label{linear}
\end{figure}

\section{Cavity QED with motion}\label{masterjunge}
In this section, we introduce the Jaynes-Cummings (J-C) model \cite{jaynes} to describe the interaction of a (moving) two-level atom with photons in a cavity mode with amplitude $u(x)=\cos(k_c x)$, where $k_c$ is the wavevector of the cavity mode as shown in figure~\ref{linear}. In the case where the atomic frequency $\omega_0$ is far detuned from the bare cavity resonance $\omega_c$, we eliminate the atomic internal degrees of freedom, to arrive at an effective optomechanical interaction between the atomic motion and cavity. We further proceed to derive an effective master equation describing the atomic motion when the cavity is externally driven by a coherent state with photon number flux $E_0^2$ and frequency $\omega_L$. We note that such a procedure would give rise to, e.g., the usual optical cooling and heating rates in a conventional optomechanical system \cite{zwerger,cooling1,cooling2}. In our case, however, we neither linearize the cavity field around a steady-state solution nor the motion, owing to the potentially large coupling between motion and the cavity field, which leads to much richer effects.

The full quantum master equation associated with the J-C model, in an interaction picture rotating with the laser frequency $\omega_L$, is given by
\begin{equation}\label{fullmodel}
 \dot{\rho} = -\I \left[H_\mathrm{JC},\rho \right] + (L_\gamma + L_\kappa)\rho \equiv L\rho.
\end{equation}
The J-C Hamiltonian including motion is given by
\begin{align}\label{ham}
 H_\mathrm{JC} = \omega_m b^\dag b - \delta_0 \sigma_\mathrm{ee} -\delta_c a^\dag a  + \sqrt{\kappa_r} E_0 (a+a^\dag)+ g_0 u(x)(a^\dag \sigma_\mathrm{ge} + h.c.).
\end{align}
It is written in terms of the detuning between laser and atom/cavity $\delta_{0/c}=\omega_L-\omega_{0/c}$, respectively, and the mechanical frequency $\omega_m$ of the external trap. Furthermore, $a$ and $b$ denote the photon and phonon annihilation operators, respectively, while $\sigma_{\mathrm{\alpha\beta}}=\ket{\alpha}\bra{\beta}$, where ${\alpha,\beta}={g,e}$ correspond to combinations of the atomic ground and excited states. $\kappa_r$ denotes the decay rate of the left cavity mirror (reflection), which also serves as the source of injection of photons. The right mirror has a decay rate of $\kappa_t$ (transmission). In addition to the external coupling, the cavity has an intrinsic loss rate $\kappa_{\mathrm{in}}$, such as through material absorption or scattering losses. The total cavity linewidth is thus $\kappa=\kappa_r + \kappa_t + \kappa_{\mathrm{in}}$. The last term of $H_\mathrm{JC}$ describes the coupling between cavity and atom with the coupling strength $g_0 u(x)$ depending on the atomic position $x = x_\mathrm{zp}( b + b^\dag)$, which can be written in terms of the zero-point motion $x_\mathrm{zp} = \sqrt{\hbar/(2 m \omega_m)}$ ($m$ being the atomic mass), and where $g_0$ is the magnitude of the vacuum Rabi splitting at the anti-node of the cavity. The Lindblad $L_c$ operator describing cavity dissipation is given by:
\begin{equation}\label{Lca}
L_\kappa \rho =- \frac{\kappa}{2} \left( a^\dag a \rho + \rho a^\dag a - 2 a \rho a^\dag \right)
\end{equation}
and the general Lindblad operator $L^{\mathrm{3D}}_{\gamma}$ for spontaneous emission into three dimensions of the atom at a rate $\gamma$ reads \cite{ritsch1}:
\begin{equation}\label{Lca1}
L^\mathrm{3D}_{\gamma}\rho =- \frac{\gamma}{2} \left( \sigma_\mathrm{ee} \rho + \rho \sigma_\mathrm{ee} -2 \int d\Omega_{\vec{u}} N_f(\vec{u}) \sigma_\mathrm{ge} e^{- \I k_c \vec{u} \cdot \vec{r}} \rho e^{\I k_c \vec{u} \cdot \vec{r}} \sigma_\mathrm{eg}\right). 
\end{equation}
This process, additionally to the emission of a photon, causes a recoil of $k = \frac{\omega_0}{c} \approx k_c$ opposite
to the direction $\vec{u}$ of the emitted photon, which is
integrated over solid angle $(d\Omega_{\vec u})$ and weighted by the distribution function $N_f(\vec u)$ corresponding to the dipole emission pattern.
However, to provide a simpler model that qualitatively captures the correct behavior, we will just consider one single direction of spontaneous emission along the positive cavity axis ($x$). With a single spontaneous emission direction we can write
\begin{equation}\label{Lca}
L_{\gamma}\rho = - \frac{\gamma}{2} \left( \sigma_\mathrm{ee} \rho + \rho \sigma_\mathrm{ee} -2 \sigma_\mathrm{ge} e^{- \I k_c  x} \rho e^{\I k_c x} \sigma_\mathrm{eg}\right). 
\end{equation}
Now we consider the dispersive regime $\Delta = \omega_0 - \omega_c \gg g_0, \kappa, \gamma$, where the atom-cavity detuning is large. Thus the single-excitation eigenstates of the J-C Hamiltonian are either mostly atomic ($\ket{\psi_+} \approx \ket{e,0}$) or photonic ($\ket{\psi_-}\approx \ket{g,1}$), where 0,1 denote the intra-cavity photon Fock state number.
These eigenstates have corresponding eigenenergies $E_1^+ \approx \omega_0 +\frac{g_0^2}{\Delta} u^2(x)$ and $E_1^- \approx \omega_c -\frac{g_0^2}{\Delta} u^2(x)$, respectively. Here, we focus on the case when the system is driven near resonantly with the photonic eigenstate. In that limit, the atom can approximately be viewed as a classical dielectric that provides a position-dependent cavity shift with an effective optomechanical coupling strength $\propto \frac{g_0^2}{\Delta}$. We will derive this effective optomechanical model now in more detail.
\subsection{Effective Optomechanical Model}\label{singlecoup}
For large laser-atom detunings $\delta_0 \gg g_0$, the atomic ground state population is approximately one, which allows for an effective elimination of the atomic excited state \cite{ritsch1,stefan} using the Nakajima-Zwanzig projection operator formalism \cite{naka,zwanzig,zwerger} (details in \ref{ap:atom}). The resulting effective master equation is given by
\begin{equation}\label{mastaop}
 \dot{\rho} = -\I \left[H_\mathrm{om},\rho \right] + L_\mathrm{om}\rho, 
\end{equation}
with an effective optomechanical Hamiltonian
\begin{equation}\label{gfn}
 H_\mathrm{om}= \omega_m b^\dag b - \Delta_c(x) a^\dag a + \sqrt{\kappa_r} E_0 (a + a^\dag).
\end{equation}
The position dependent cavity-laser detuning is given by
\begin{equation}\label{cav}
 \Delta_c(x)=  \delta_c -  \frac{g_0^2 \delta_0}{\delta_0^2 + \frac{\gamma^2}{4}} u^2(x), 
\end{equation}
which now accounts for the cavity shift arising from off-resonant coupling to the atomic transition. The system losses are given by the effective Liouvillian
\begin{align}\label{Lc}
L_\mathrm{om}\rho & \nonumber =- \frac{\kappa}{2} \left( a^\dag a \rho + \rho a^\dag a - 2 a \rho a^\dag \right)  \\ & -  \frac{\gamma}{2}\frac{ g_0^2}{ \delta_0^2 + \frac{\gamma^2}{4}} \left(u^2(x) a^\dag a \rho + \rho a^\dag a u^2(x) -2 a u(x) e^{- \I k_c x} \rho e^{\I k_c x} u(x) a^\dag\right), 
\end{align}
which describes the broadening of the cavity linewidth due to atomic spontaneous emission,  
\begin{equation}\label{ka}
 \kappa(x) = \kappa + \gamma  \frac{g_0^2}{\delta_0^2+ \frac{\gamma^2}{4}} u^2(x). 
\end{equation}
Aside from \ref{ap:full}, where we discuss in greater detail the corrections to and limitations of the effective model, we will work in regimes where the atomic contribution is negligible compared to the (large) bare cavity linewidth.

In typical treatments of optomechanical systems, the position-dependent shift in equation~(\ref{cav}) would only be treated to linear order in the displacement, with the justification that the maximum possible displacement is very small. However, for atoms, the zero-point motion can be comparable to the optical wavelength (the scale over which $u(x)$ varies), a ratio that can be characterized by the Lamb-Dicke parameter $\eta_{LD}\equiv k_c x_\mathrm{zp}$. For example, taking  a recoil frequency $\omega_\mathrm{rec}=2 \pi \times 6.8 \,\mathrm{kHz}$ corresponding to $^{40}\mathrm{Ca}^+$-ions and a trap frequency of $\omega_m = 2 \pi \times 0.1\, \mathrm{MHz}$ results in $\eta_\mathrm{LD} = \sqrt{\omega_\mathrm{rec}/\omega_m} \approx 0.26$. For $\eta_{LD}\sim 1$, the atomic wavepacket would have significant weight both in a cavity anti-node and node, with an associated cavity frequency shift of
\begin{equation}\label{gom}
 g_\mathrm{om} =  - \frac{g_0^2 \delta_0}{\delta_0^2 + \frac{\gamma^2}{4}}
\end{equation}
 and zero, respectively. As our perturbative treatment is valid for $\delta_0\gtrsim g_0$ (see \ref{ap:full}), one sees that strong optomechanical coupling $g_\mathrm{om}\gtrsim \kappa$ can be achieved if the strong coupling regime of conventional cavity QED ($g_0>\kappa$) is realized. The standard optomechanical Hamiltonian (linearized in displacement) describing interactions between single-photons and single-phonons is given by $H_\mathrm{oms} = g_m (b^\dag + b)a^\dag a$, where $g_m = \Delta'_c(x_0) x_\mathrm{zp} \sim g_\mathrm{om} \eta_\mathrm{LD}$. Thus, in order to achieve strong optomechanical coupling on the single-photon, single-phonon level ($g_m \gtrsim \kappa$), additionally a sufficiently large Lamb-Dicke parameter $\eta_\mathrm{LD}$ is required.
 Given the above considerations, we next derive an effective master equation for the atomic motion alone that is valid for strong and nonlinear optomechanical coupling, which can be viewed as a generalization of the typical optically-induced cooling and heating rates obtained for linearized optomechanical coupling
 \cite{zwerger, cooling1, cooling2}.
 Our master equation also complements previous work investigating intra-cavity optical forces on atoms in the semi-classical limit \cite{othercooling1,othercooling2,ritsch1,othercooling4, gio}.
 \subsection{Effective Master Equation for Motion}
Starting with equation~(\ref{mastaop}) we can use the Nakajima-Zwanzig technique to effectively eliminate the cavity degrees of freedom (\ref{ap:cav}). Here, for simplicity we assume that spontaneous emission can be ignored. The resulting master equation for atomic motion in conventional Lindblad-form is then given by:
\begin{equation}\label{nmastacon}
 \dot{\rho} = -\I [H_{\mathrm{m}},\rho] - \frac{1}{2}\left( J^\dag J \rho + \rho J^\dag J \right) + J\rho J^\dag . 
\end{equation}
The Hermitian Hamiltonian and jump operators are given respectively by
\begin{equation}
H_\mathrm{m} = \omega_m b^\dag b +\frac{\kappa_r E_0^2  \Delta_c(x)}{\Delta_c^2(x) + \frac{\kappa^2}{4}}
\end{equation}
and 
\begin{equation}\label{jj}
 J =  \frac{\I \sqrt{\kappa \kappa_r} E_0 }{\Delta_c(x) + \I \frac{\kappa}{2}}.
\end{equation}
We will provide an intuitive picture of this master equation in section \ref{sec:con}. Now we focus on the effective mechanical potential which arises in the Hamiltonian. We can always rewrite a master equation in terms of an effective non-Hermitian Hamiltonian $H_c$ which then contains a complex potential:
\begin{equation}\label{nmastaef}
 \dot{\rho} = -\I (H_c \rho - \rho H_c^{\dagger}) + J\rho J^\dag 
\end{equation}
\begin{equation}
H_\mathrm{c} = \omega_m b^\dag b + V(x)
\end{equation}
with
\begin{equation}\label{po}
 V(x) = \frac{\kappa_r E_0^2  \Delta_c(x)}{\Delta_c^2(x) + \frac{\kappa^2}{4}} - \frac{\I}{2} J^\dag J.
\end{equation}
\begin{figure} 
\centering
\includegraphics[width=0.7\textwidth]{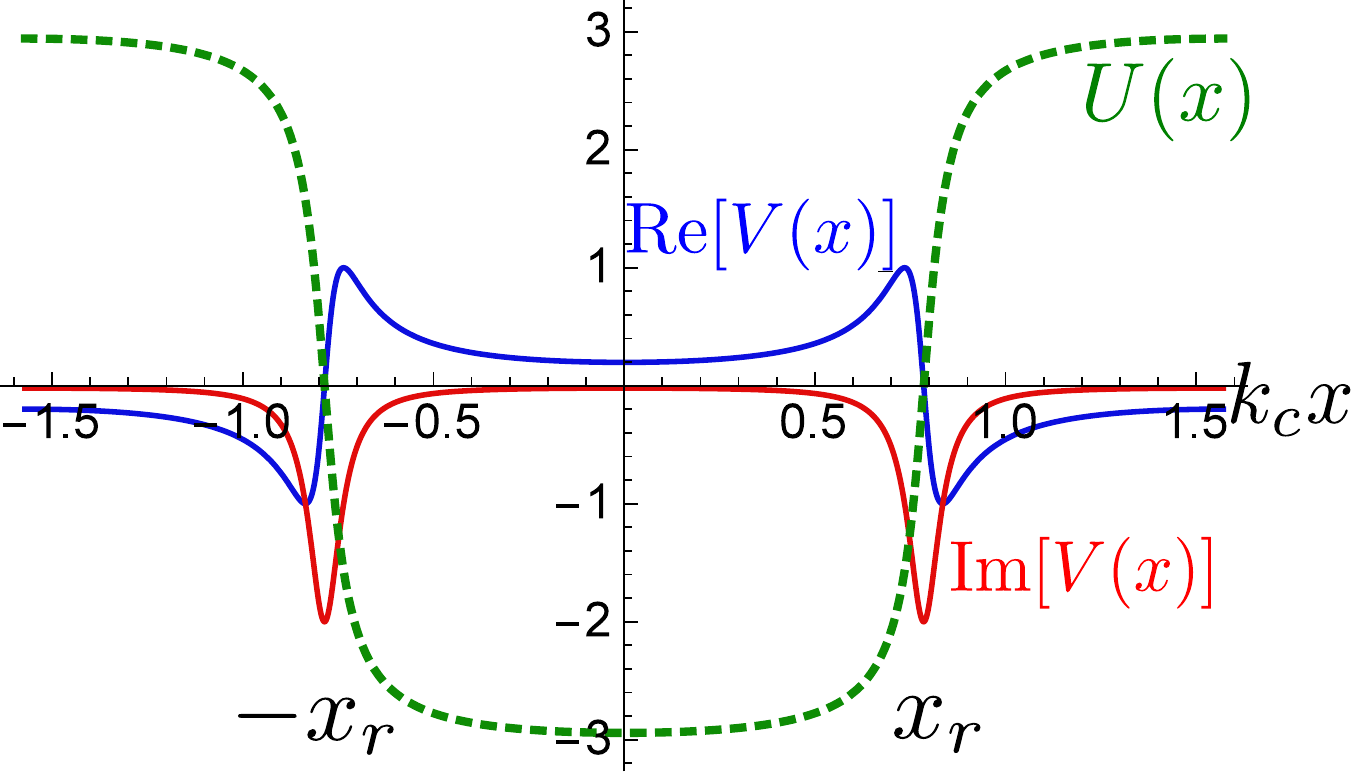} 
\caption{ \textbf{Quantum and classical mechanical potential arising from a coherently driven cavity mode} \\
 Real part  $\mathrm{Re}[V(x)]$ (blue) and imaginary part $\mathrm{Im}[V(x)]$ (red) of the quantum potential equation~(\ref{po}) as a function of position. Also plotted is the classical potential $U(x)$ (dahsed, green) derived by integrating the expectation value of the force acting on the atom. One can observe that the real part of the quantum potential is significantly different from the classical expectation value.
  Here, we choose a laser frequency $\omega_L$ such that the resonant position $k_c x_r = \pi/4$, and Jaynes-Cummings parameters of $g_0/\kappa\sim 20$ and $\delta_0=-2g_0$ (yielding an effective optomechanical coupling strength of $g_\mathrm{om} \sim 10 \kappa$). 
  The potentials are plotted in units of $\hbar (\kappa_r/\kappa) E_0^2$.}
\label{poten}
\end{figure}
The real and imaginary parts of the complex potential $V(x)$ are illustrated in figure~\ref{poten}.  As the resonance frequency of the cavity depends on the position of the atom, there can be atomic positions for which the cavity is resonant with the coherent drive. These positions $x_r$ are called resonant positions and are defined by $\Delta_c(x_r) = 0$. Around these positions, the real part of the potential changes sign and the imaginary part has sinks indicating increased heating around those positions.

It is also interesting to compare the ``coherent'' potential, $\mathrm{Re}[V(x)]$, with the classical potential $U(x)$ as derived from the average force $F(x)=d \langle p \rangle /dt=\mathrm{Tr}(p \rho)$ on the atom, and defined via $dU/dx=-F(x)$. The result is given by 
\begin{equation}
 U(x) = -2 \frac{\kappa_r}{\kappa} E_0^2 \arctan\left(\frac{2 \Delta_c(x)}{\kappa}\right),
\end{equation} 
which agrees with a previous, completely classical analysis of a dielectric object trapped in a cavity  \cite{ich}.
The potential is illustrated in figure~\ref{poten}. For large $g_\mathrm{om}/\kappa$, $U(x)$ is seen to approach a square well, with the walls of the well aligning with the resonant positions $\sim x_r$ where the large intracavity field results in a large classical restoring force. By comparing $V(x)$ and $U(x)$, it is clear that a significant contribution of the average force must arise from the stochastic process associated with the quantum jumps $J$. As one consequence, although it would be highly interesting to realize a square well for atoms (leading, e.g., to a highly anharmonic phonon spectrum), the direct quantization of $U(x)$ in this case is not meaningful.
\section{Single-photon scattering theory: Optomechanical strong coupling with unresolved sidebands} \label{scattar}
A complementary physical picture of the optomechanical coupling between an atom and cavity can be gained by considering not a coherent external drive, but single incident photons. From equation~(\ref{mastaop}), the effective non-Hermitian Hamiltonian associated with an undriven system is
\begin{equation}\label{int}
 H_{\mathrm{eff}} =  \omega_m b^{\dagger}b -(\Delta_c(x) + \I \frac{\kappa}{2}) a^\dagger a   
\end{equation} 
where $\Delta_c(x) = \omega_L - \omega_c(x)$ is the position-dependent detuning between photon frequency $\omega_L$ and cavity frequency $\omega_c(x)= \omega_c - g_\mathrm{om} u^2(x)$. To be specific, we will consider single photons incident through the left mirror (see figure~\ref{linear}), which has a decay rate back into the reflection channel of $\kappa_r$. The right mirror is coupled to the controlled transmission channel with $\kappa_t$.  The total cavity linewidth is thus $\kappa=\kappa_r + \kappa_t$. For simplicity we ignore here an intrinsic loss rate, although it is straightforward to include later on.

A connection can be made between the eigenstates of $H_\mathrm{eff}$ and the properties of single-photon scattering via the S-matrix formalism. Formally, the S-matrix describes a coherent evolution mapping an input state ($t= - \infty$) to an output state ($t= + \infty$):
\begin{equation}\label{ss}
 \ket{\Psi_\mathrm{out}(\omega_L)} = S\ket{\Psi_\mathrm{in}(\omega_L)}. 
\end{equation}
Here, we assume a single monochromatic photon with frequency $\omega_L$ incident on the left cavity mirror
\begin{equation}\label{in} 
 \ket{\Psi_\mathrm{in}(\omega_L)}=\ket{(\omega_L)_\mathrm{left},0},
\end{equation}
whereas the optomechanical system initially is in its ground state represented by the second entry in the ket state.
Generically the output state will consist of a superposition of $n$ phonons in the mechanical state, which were excited by the incoming photon, and an outgoing photon of energy $\omega_L-n\omega_m$ in either the reflection port (r) or transmission port (t):
\begin{equation}\label{out}
 \ket{\Psi_{\mathrm{out}}(\omega_L)} = \sum_n S_\mathrm{{r,n}}(\omega_L) \ket{(\omega_{\mathrm{L}}- n\omega_m)_r,n} + \sum_n S_{\mathrm{t},n}(\omega_L) \ket{(\omega_{\mathrm{L}}- n\omega_m)_t,n}.
\end{equation}
 Due to a connection between the scattering matrix and the Heisenberg input-output operators \cite{shan} one can express the S-matrix elements in terms of the eigenvalues $\lambda_\beta$ and eigenstates $\ket{\beta}$ of the effective Hamiltonian $H_\mathrm{eff}$ \cite{marco}.
We provide a detailed derivation of the S-matrix elements in \ref{ap:scat}.
In reflection, the output consists of a superposition between a non-interacting propagating photon ($\delta_{n,0}$) and photon emission from the excited optomechanical system: 
\begin{equation}\label{smex}
 S_{\mathrm{r,n}}(\omega_{L}) = \delta_{n,0} + \I \kappa_r \sum_{\beta} \langle 1_c,n|\beta \rangle \frac{1}{\lambda_\beta} \langle \beta|1_c,0 \rangle.
\end{equation}
Here, $\langle 1_c,n|\beta \rangle $ is the projection of the eigenstates $\ket{\beta}$ onto the basis states $\bra{1_c,n}$ with $1_c$ referring to a single photon inside the cavity mode.
Similarly, the matrix elements for photon transmission are given by
\begin{equation}\label{sm0} 
 S_{\mathrm{t,n}}(\omega_L) =  \I \sqrt{\kappa_\mathrm{t} \kappa_r} \sum_{\beta} \langle 1_c,n|\beta \rangle \frac{1}{\lambda_\beta} \langle \beta|1_c,0 \rangle.
\end{equation}
The matrix element for photon transmission lacks the contribution from the non-interacting propagating photon as the input channel on the transmitting side of the cavity is in the vacuum state. To proceed further, we assume in the following that a detector cannot effectively resolve the frequency of the outgoing photon. 
Then, we can effectively write the outgoing state as
\begin{equation}\label{state} 
 \ket{\Psi_{\mathrm{out}}(\omega_L)} = S_{\mathrm{r}}(\omega_L,x) \Psi_0(x) \ket{1_r} +  S_{\mathrm{t}}(\omega_L,x) \Psi_0(x) \ket{1_\mathrm{t}},
\end{equation}
where $\ket{1_{r/t}}$ indicates an outgoing reflected/transmitted photon, respectively, and $\Psi_0(x)$ is the initial motional wave function of the atom. The entanglement between the photon frequency and the motional state has been suppressed, as we have assumed that any projective measurement of a photon in either port is not frequency-resolving. Furthermore, we now assume that we operate in the sideband-unresolved limit $\kappa\gg\omega_m$. The Hamiltonian $H_\mathrm{eff}$ is approximately diagonal in the position basis, as the optomechanical interaction dominates over the free Hamilontian $\omega_m b^\dag b$ in $H_\mathrm{eff}$ (equation~\ref{int}). Thus, the eigenvalues of $H_\mathrm{eff}$ are approximately $\lambda \approx -\Delta_c(x)- \I \frac{\kappa}{2} $ and the scattering matrix elements can be simply written as
\begin{equation}\label{sex1}
  S_r(\omega_L,x) =  1 -   \frac{\I \kappa_r}{\Delta_c(x) + \I \frac{\kappa}{2}}
\end{equation}
and 
\begin{equation}\label{s01}
  S_{\mathrm{t}}(\omega_L,x) =  -   \frac{\I\sqrt{\kappa_\mathrm{t} \kappa_r} }{\Delta_c(x) + \I \frac{\kappa}{2}}.
\end{equation}
As the shape of the mechanical wave function  after the decay of a single photon into one specific channel is the product between the corresponding S-matrix element and the initial wave function $\Psi_0(x)$, we observe that the shape of the mechanical wave function after one such scattering event is strongly entangled with whether the decaying photon is reflected or transmitted. 

\label{sec:con}
Motivated by the observation that the scattering matrices $S_t$ and $S_r$ of Eqs. (\ref{sex1}) and (\ref{s01}) are very similar to the jump operators $J$ (equation~\ref{jj}), we express the master equation (\ref{nmastacon}) in a way that its jump operators correspond to the single photon scattering matrices:
\begin{equation}\label{ma}
 \dot{\rho} = -\I (H_s\rho- \rho H_s^\dagger) + E_0^2 (S_r\rho S_r^\dag + S_t\rho S_t^\dag) 
\end{equation}
with the Hamiltonian
\begin{equation}\label{he}
 H_s = \omega_m b^\dag b - \frac{\I}{2} E_0^2.
\end{equation}
Written in this form the connection between scattering theory and jump formalism becomes clear. The non-Hermitian term in $H_s$ describes the rate that quantum jumps are applied to the motional wave function, which corresponds to the rate $E_0^2$ of incident photons on the cavity. The jump operators themselves, $J_\mathrm{r/t}=E_0 S_\mathrm{r/t}$, with ($J_r^{\dagger}J_r+J_t^{\dagger}J_t=E_0^2$), are proportional to the single-photon scattering matrix elements in reflection and transmission, encoding the two processes by which the original wave function can change by becoming entangled with a scattered photon.
Interestingly, the coherent part of the potential, $\mathrm{Re}[V(x)]$ in equation~(\ref{po}), is seen to arise from the term  $S_r \rho S_r^{\dagger}$ in equation~(\ref{ma}), and specifically from the interference between the incident and scattered components (first and second terms on the right of equation~(\ref{sex1}), respectively).
\section{Quantum Effects due to Zero-Point motion} 
We have already seen that the scattering of a single photon on a cavity containing an atom leads to an entangled output state (\ref{state}). This output state describes the coexistence of the possibilities of photon reflection and photon transmission and how the wave function of the atom gets modified for each of those events.  
 We now proceed to describe some of the relevant observational consequences.
 \begin{figure}
\centering
\includegraphics[width=0.85\textwidth]{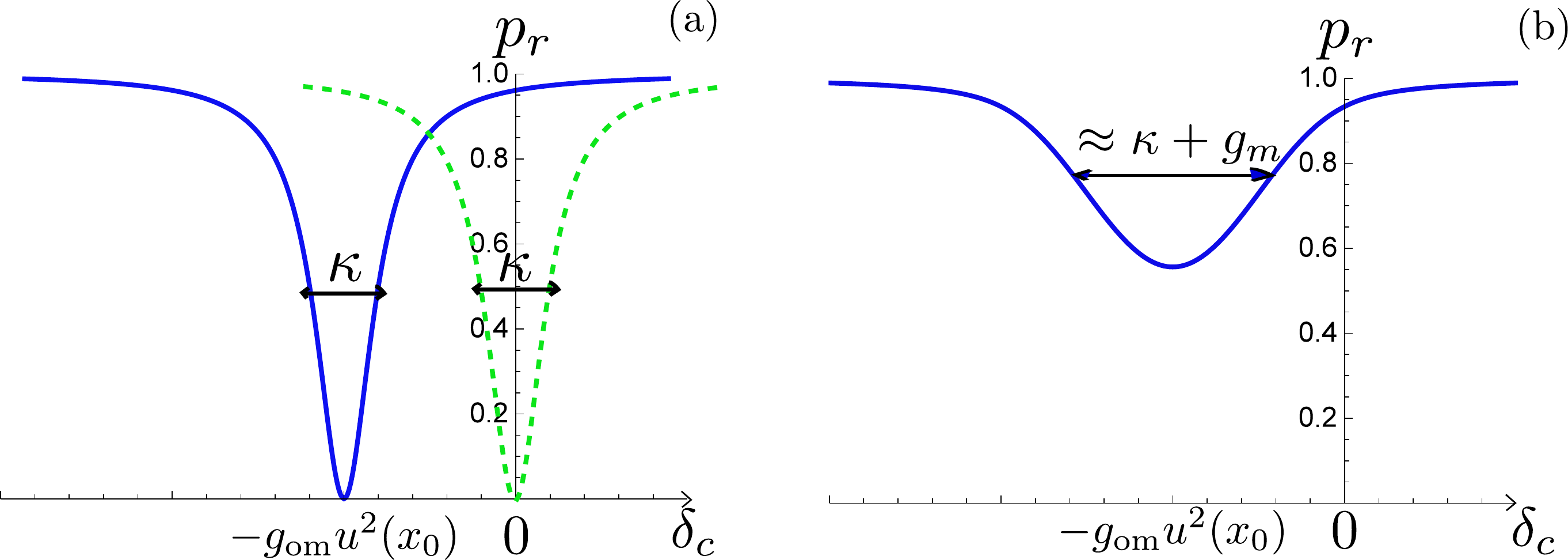} 
\caption{\textbf{Reflection spectrum $p_r$ as a function of laser frequency $\omega_L$.}
Here, we take critical coupling ($\kappa_r=\kappa/2$) and a trap equilibrium position of $k_c x_0=\pi/4$. We assume the initial atomic wave function is in the motional ground state. \\
 \textbf{a)} If the zero-point motion is unresolved, the reflection spectrum (blue) just behaves like the reflection spectrum of an empty cavity (green, dashed) but is shifted to a new resonance $\omega_c(x_0)$. Here we choose $g_\mathrm{om} = \kappa$ and $\eta_\mathrm{LD} = 0.01$, implying $r_\mathrm{zp} = 0.02$.  \\
\textbf{b)} If the zero-point motion is resolved, the reflection spectrum is broadened by roughly $g_m$ and becomes shallower. 
Here we choose $g_\mathrm{om} = 5 \kappa$ and $\eta_\mathrm{LD}=0.2$, implying $r_\mathrm{zp} = 2$.}
\label{JCeffs}
\end{figure} 

We can expand the position-dependent cavity detuning
around a resonant position $x_r$ (defined by $\Delta_c(x_r) = 0$) until linear order:
\begin{equation}\label{exp}
 \Delta_c(x) \approx \delta_c + g_\mathrm{om} u^2(x_r) - g_\mathrm{om} \sin(2 k_c x_r) k_c (x-x_r).
\end{equation}
This is a good approximation in the Lamb-Dicke regime $\eta_\mathrm{LD} \ll 1$. In order to predict observables, linearizing displacement is also a good approximation for $g_\mathrm{om} \gg \kappa$, even if $\eta_\mathrm{LD} \sim 1$, since then the cavity frequency shifts out of resonance for displacements $k_c \delta x \ll 1$. The term $\sin(2 k_c x_r)$ indicates that the cavity frequency is most sensitive to displacements if $k_c x_r = \pm \pi/4$, halfway between a cavity node and anti-node. 
 \begin{figure}
\centering
\includegraphics[width=\textwidth]{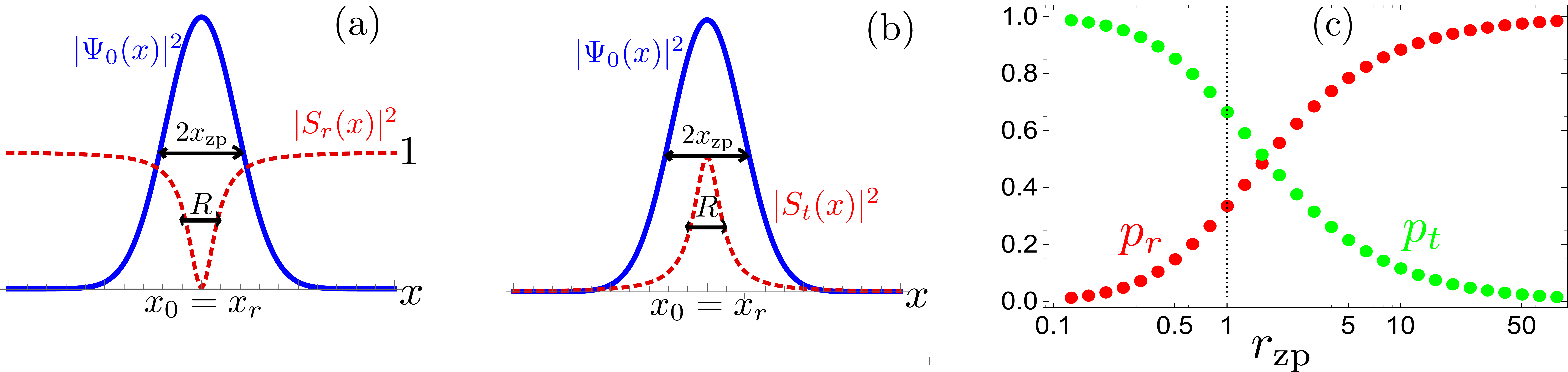} 
\caption{\textbf{Resolution beyond zero-point uncertainty} \\
 \textbf{a)}  For $r_\mathrm{zp} = 2$, the spatial width $\sim 2 x_\mathrm{zp}$ of the atomic probability density $|\Psi_0(x)|^2$ (blue) exceeds the spatial resolution $R$, which corresponds to the width of the absolute value of the scattering 
matrix $|S_r(x)|^2$ (red dashed). As the cavity is only resonant with an incoming photon if the atom is located within $R$, there is a large probability that the cavity is off-resonant, even though $\omega_L = \omega_c(x_0)$. The probability of reflection is calculated by the overlap of both plotted functions.   \\
\textbf{b)} Same as a), but the absolute value of the S-matrix for transmission (red, dashed). \\
 \textbf{c)} Probability of photon reflection $p_r$ (red) and  transmission $p_t$  (green) as a function of zero-point resolution $r_\mathrm{zp}$, for an incident photon that is resonant with the cavity in the limit that atomic motion fluctuations are ignored (i.e., $\delta_c=-g_\mathrm{om}u^2(x_0))$. One sees that for large $r_\mathrm{zp}$, the probability of transmission becomes negligible, because the probability of finding the atom within $R$ (which would imply a resonant system and consequent transmission) approaches zero for $r_\mathrm{zp} \gg 1$.} 
\label{intuitive}
\end{figure}
Then it can be seen that if the atomic wave function is centered around $k_c x_0 = k_c x_r = \pi/4$, the cavity frequency shifts by a linewidth $\kappa$, if the atom moves a distance of $k_c R = \kappa/g_\mathrm{om}$. As the transmission/reflection of a single, near-resonant photon changes significantly as its frequency varies over a cavity linewidth, $R$ can be viewed as the spatial resolution over which the single photon "learns" about the atomic position via its scattering direction.
We will now define the zero-point resolution
\begin{equation}
 r_\mathrm{zp} \equiv (2 x_\mathrm{zp})/R=(2 g_m)/\kappa,
\end{equation}
with $g_m = g_\mathrm{om} \eta_\mathrm{LD}$ being the single-photon, single phonon coupling strength as defined in section \ref{singlecoup}. 
The zero-point resolution tells us how much finer the resolution of an incident photon is compared to the width of the atomic wave function. It distinguishes two regimes: unresolved zero-point motion $r_\mathrm{zp} \ll 1$, which corresponds to the usual regime of weak optomechanical interactions, and the resolved zero-point motion regime $r_\mathrm{zp} \gg 1$, where the resolution of the system becomes smaller then the zero-point motion, which is until now unexplored and which gives rise to novel effects as we will demonstrate in the following. 
 \subsection{Influence of the zero-point motion on the reflection spectrum} \label{zpo}
 Here, we assume the atom to be initially in its motional ground state $\Psi_0(x) \propto e^{- \frac{1}{4} (x-x_0)^2/x_\mathrm{zp}^2}$ with 
a trap equilibrium $k_c x_0 = \pi/4$ and $\kappa_r=\kappa/2$ (critical coupling). The spectrum of reflection, as a function of the incident photon frequency $\omega_L$, is then given by
 \begin{equation}\label{refl}
 p_\mathrm{r}(\omega_L) = \int dx |S_r(\omega_L,x)|^2 |\Psi_0(x)|^2.
\end{equation}
 Figure~\ref{JCeffs}(a) shows $p_r$ as a function of cavity detuning $\delta_c = \omega_L-\omega_c$ for $r_\mathrm{zp}\ll 1$ (unresolved zero-point motion). The green dashed line is the reflection spectrum of an empty cavity with decay rate $\kappa$. The blue solid line is calculated with equation~(\ref{refl}) for $r_\mathrm{zp} =0.02$, where $p_r \approx |S_r(\omega_L,x_0)|^2$. One can see that it exhibits the same Lorentzian response as an empty cavity, but with a resonance frequency
 shifted by $-g_\mathrm{om} u^2(x_0)$. Figure~\ref{JCeffs}(b) shows the reflection spectrum $p_r$ for resolved zero-point motion $r_\mathrm{zp} = 2$.
We observe that the probability of reflection is strongly increased for $\delta_c = - g_\mathrm{om} u^2(x_0)$, compared to the case of small $r_\mathrm{zp}$.
This behavior can be understood from equation~(\ref{exp}). In particular, the resonance frequency of the coupled atom-cavity system depends on the position of the atom, and $\delta_c=-g_\mathrm{om} u^2(x_0)$ corresponds to the resonance of the most likely atomic position. However, the large spread of the atomic wave function results in a large uncertainty of the resonance frequency, which increases the reflection probability. Conversely, an incident photon with frequency far from $\delta_c=-g_\mathrm{om} u^2(x_0)$ sees a decreased reflection probability (thus the broadening of the spectrum), as there is some chance that the spread in atomic position allows the coupled system to be on resonance with the photon. This is illustrated in figure~\ref{intuitive}(a), where we plot the atomic probability density $|\Psi_0(x)|^2$ (blue) and the absolute value of the reflection S-matrix $|S_r(x)|^2$ (red dashed)  (equation~\ref{sex1}) as a function of position $x$ and for $r_\mathrm{zp}=2$. One can see, that the width of the atomic wave function $\sim 2 x_\mathrm{zp}$ exceeds the spatial resolution $R$, within which the cavity is resonant.
 For completeness, we also provide a plot of the absolute value of the transmission S-matrix $|S_t(x)|^2$ (red dashed) in figure~\ref{intuitive}(b). 
 Figure~\ref{intuitive}(c) shows the probability of reflection and transmission for $\delta_c = - g_\mathrm{om} u^2(x_0)$ as a function of $r_\mathrm{zp}$. For $r_\mathrm{zp}\ll 1$ the probability of reflection vanishes and the transmission approaches unity as it would for an empty resonant cavity. However, with increasing $r_\mathrm{zp}$ it becomes less likely to find the atom within the spatial resolution $R$ within which the cavity is resonant, leading to an increase of $p_r$. Finally, the reflection probability $p_r$ approaches unity for $r_\mathrm{zp}\gg 1$. 
 
 Most of this plot is already   
experimentally accessible with current technology. For example a neutral atom trapped in its ground state inside photonic crystal cavities can reach $r_\mathrm{zp} \sim 10$ (\ref{ap:cryst}.1) whereas a current fiber cavity experiment reaches $r_\mathrm{zp} \sim 1$ (\ref{ap:cryst}.2). While measuring $p_r$, the zero-point resolution $r_\mathrm{zp}$ can then be gradually decreased by increasing the atom-cavity detuning $\omega_0-\omega_c$, increasing trap frequency $\omega_m$ or by moving the trap equilibrium $x_0$ away from the position of maximal optomechanical coupling $k_c x_0= \pm \pi/4$. This procedure would experimentally reproduce parts of figure~\ref{intuitive}(c).    
\subsection{Entanglement and conditional projection of the atomic wave function}\label{ent}
Having previously investigated the unconditional reflection spectrum of an incident photon, we now study more carefully the correlations that build up between the atomic motion and photon reflection or transmission for the case when the trap equilibrium falls at the resonant position ($x_0=x_r$).  As the atom is in a coherent superposition of being within the spatial resolution $R$ and not, and an incoming photon gets transmitted if the atom is within that spatial resolution and reflected if otherwise, the resulting state (equation~\ref{state}) is entangled. 
 \begin{figure}
\centering
\includegraphics[width=0.85\textwidth]{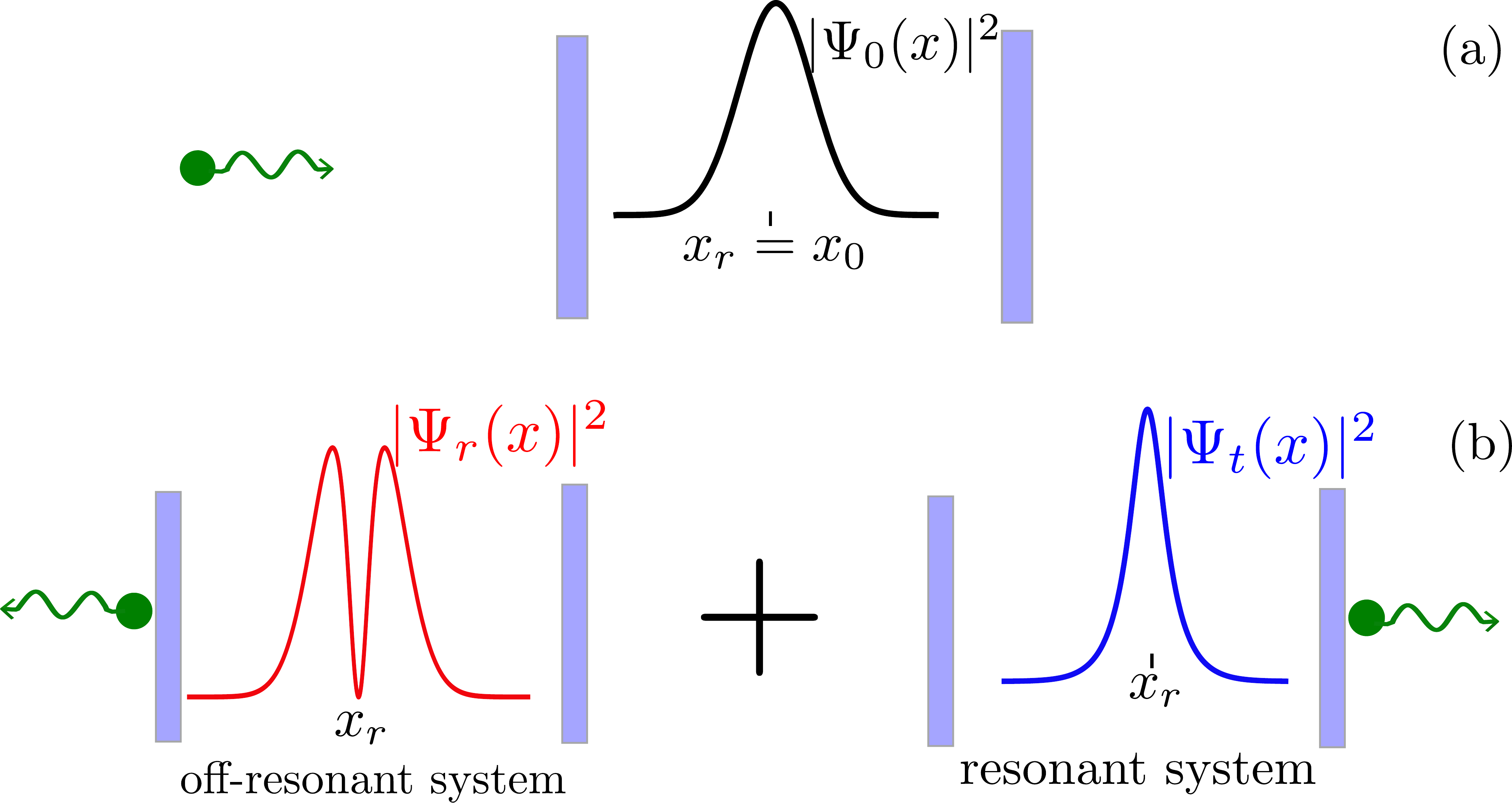} 
\caption{\textbf{Illustration of a single photon scattering event for resolved zero-point motion} \\
 \textbf{a) Input state:} An incident photon (green) with a frequency ensuring $x_0=x_r$ is flying towards a cavity containing a trapped atom with probabilitiy density $|\Psi_0(x)|^2$ (black). Due to its zero-point uncertainty, the system is in an effective superposition of resonance frequencies.
 This input state is given by equation~(\ref{in}). \\
\textbf{b) Output state:} Illustration of the entangled output state given by equation~(\ref{state}), which is a superposition of the photon being reflected, which implies an off-resonant system and a photon being transmitted, which implies a resonant system. The plotted probability densities $|\Psi_\mathrm{r/t}(x)|^2$ are the normalized product of $|\Psi_0(x)|^2$ and the respective scattering matrix $|S_\mathrm{r/t}(x)|^2$ of figure~\ref{JCeffs}(a) and \ref{JCeffs}(b), where $r_\mathrm{zp}=2$.  For this value, the probability of reflection is $p_r \approx 0.56$.
} 
\label{entangled}
\end{figure}
Given that the photon has been transmitted, the normalized conditional wave function is given by
\begin{equation}\label{psit}
 \Psi_t(x) = p_t^{-1/2} S_t(x) \Psi_0(x).
\end{equation}
Its probability density is propotional to the product of $|\Psi_0(x)|^2$ and $|S_\mathrm{t}(x)|^2$ as individually drawn in figure~\ref{intuitive}(b). Thus, for $r_\mathrm{zp} \gg 1$, the transmission of a photon projects the atom into a narrow spatial region $\Delta x \sim 1/R$ around the resonant position, which is consistent with the photon having seen a resonant cavity response.

In contrast, the reflection of a photon projects the atom away from that same spatial region, which results in a hole around $x_r$ with width $\Delta x \sim 1/R$. This is consistent with the photon having seen an off-resonant cavity. The normalized conditional wave function after a photon reflection is then given by
\begin{equation}\label{psir}
 \Psi_r(x) = p_r^{-1/2} S_r(x) \Psi_0(x).
\end{equation}
As individually drawn in figure~\ref{intuitive}(a), its probability density is propotional to the product of $|\Psi_0(x)|^2$ and $|S_\mathrm{r}(x)|^2$. Figure~\ref{entangled}(a) shows an illustration of the unentangled input state. 
The atom (black) is in its motional ground state, centered around $x_0=x_r$, while a single photon (green) is incident and resonant with the atom-cavity system for this position. 
In  figure~\ref{entangled}(b) we illustrate the entangled output state for $r_\mathrm{zp}=2$. We illustrate how the transmission or reflection of a photon are entangled with atomic wave functions $\Psi_t(x)$ or $\Psi_r(x)$ consistent with the respective scattering process, for the same parameters as in figures~\ref{JCeffs}(a) and (b). 

Interestingly, in the unresolved zero-point motion regime $r_\mathrm{zp}\ll1$ the scattering matrix for reflection is proportional to $x$: $S_r(x) \approx - 2 \I x/R$. This leads to a final conditional wave function  $\Psi_r(x) \propto x \Psi_0(x)$ which corresponds to a single-phonon Fock state. 
This represents the high-fidelity generation of a single-phonon Fock state, which is heralded on detection of a reflected photon (the probability of a single photon being reflected itself is quite low, $p_r\approx r^2_\mathrm{zp}$). This approach is distinct from previous proposals for heralded generation, involving the detection of a Stokes-scattered photon in the sideband resolved regime \cite{kip}. 

\begin{figure}
\centering
\includegraphics[width=\textwidth]{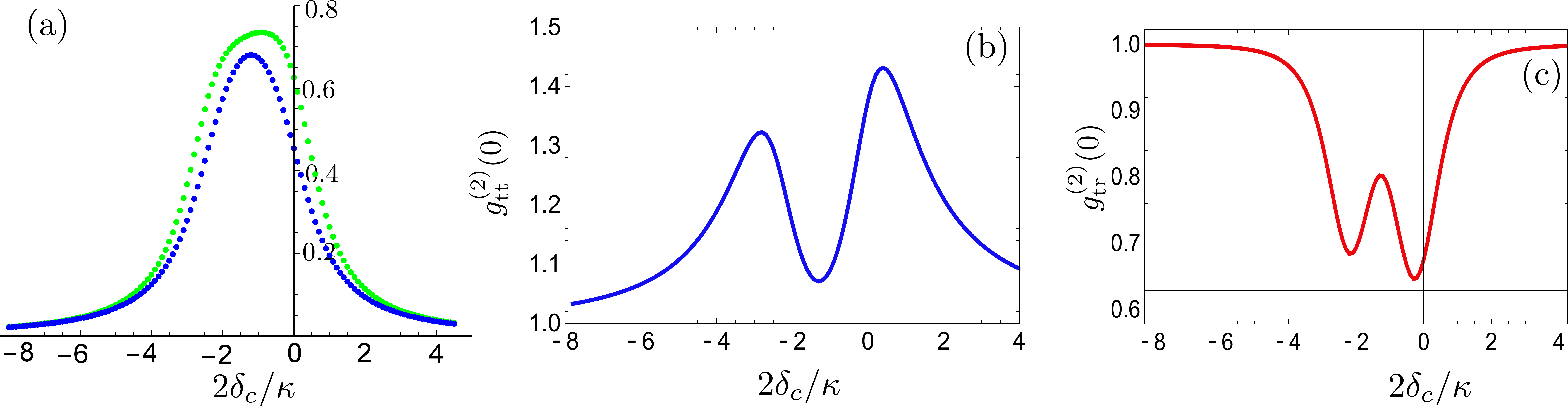} 
\caption{\textbf{Photon statistics due to wave function projection} for a fixed trapping position $k_c x_0 = \pi/4$.  \\ \textbf{(a)} Probability of photon transmission $p_t$ (blue) as a function of cavity detuning $\delta_c$  and the conditional probability of transmission, given that a photon just has been transmitted $p(t|t)$ (green). We observe that a transmitted photon increases the probability of transmitting again.
 \\
 \textbf{(b)} The second-order correlation function $g_\mathrm{tt}^{(2)}(0)$ of the transmitted field as a function of $\delta_c$ shows bunching due to the increased likelihood of detecting a transmitted photon after the transmission of a first photon.
 \\
 \textbf{(c)}  The second-order cross-correlations $g_\mathrm{tr}^{(2)}(0)$ between the transmitted and the reflected field as a function of $\delta_c$ shows anti-bunching. \\
Here we use parameters of an existing fiber cavity QED experiment with trapped $^{40}\mathrm{Ca}^+$-ions with recoil frequency $\omega_\mathrm{rec}= 2 \pi \times 6.8 \, \mathrm{kHz}$, see \ref{ap:cryst}.2(i). The parameters are $g_0 = 2 \pi \times 41 \, \mathrm{MHz}$, $\gamma = 2 \pi \times 11.2 \, \mathrm{MHz}$, $\kappa = 2 \pi \times 8 \, \mathrm{MHz}$. We chose $\omega_m = 2 \pi \times 50 \,\mathrm{kHz}$ and $\omega_0-\omega_c = 4 g_0$. These values correspond to a zero-point resolution of $r_\mathrm{zp}\approx 0.89$ for $2 \delta_c/\kappa \approx -1.2$ (calulated with equation (\ref{rzp})).
}
\label{fig4}
\end{figure}

The wave function after a transmission/reflection event adjusts in a way that it increases the probability of a subsequent transmission/reflection. To demonstrate this, we calculate the conditional probability of photon transmission given that a photon has just been transmitted:
\begin{equation}
 p(t|t) = \frac{1}{p_t}\int dx |S_t(x,\omega_L)|^4|\Psi_0(x)|^2.
\end{equation}
Figure~\ref{fig4}(a) shows $p(t|t)$ (green) as a function of cavity detuning $\delta_c$ for a fixed trapping position $k_c x_0 = \pi/4$.
We plot the corresponding probability of transmission $p_t$ (blue) as well, which is seen to be lower than the conditional probability. 
We use parameters of an existing fiber cavity QED experiment with trapped $^{40}\mathrm{Ca}^+$-ions (\ref{ap:cryst}.2(ii)). We chose $\omega_m = 2 \pi \times 50 \,\mathrm{kHz}$ and $\omega_0-\omega_c = 4 g_0$. The asymmetry of $p(t|t))$ is due to the nonnegligible dependence of $g_\mathrm{om}$ (equation~(\ref{gom})) and $\kappa(x_0)$ (equation~(\ref{ka})) on the laser frequency $\omega_L$ (and thus $\delta_c$) for those parameters. For $2 \delta_c/\kappa = 
- (2 g_\mathrm{om}/\kappa) u^2(x_0) \approx -1.2$ (which implies $x_r = x_0$) a zero-point resolution of $r_\mathrm{zp}\approx 0.89$ is obtained, which needs to be calculated with equation~(\ref{rzp}) as here spontaneous emission cannot be neglected. As one consequence of the higher likelihood of conditional transmission, the second-order correlation function $g_\mathrm{tt}^{(2)}(0)=\frac{1}{p_t^2}\int dx |S_t(x,\omega_L)|^4|\Psi_0(x)|^2$ of the transmitted field, given a weak coherent input state, would exhibit bunching, as shown in figure~\ref{fig4}(b). Likewise, as reflection of a first photon suppresses the probability of transmitting a second photon (and vice versa), second-order cross-correlations $g_\mathrm{rt}^{(2)}(0) = \frac{1}{p_t p_r}\int dx |S_t(x,\omega_L)|^2 |S_r(x,\omega_L)|^2 |\Psi_0(x)|^2$ between the reflected and transmitted field would exhibit anti-bunching (figure~\ref{fig4}(c)).

\subsection{Motional heating induced by entanglement}
Each projection of the atomic wave function is associated with an increase in energy. We will now show that this energy can vastly exceed the energy added in free space or in a trap. In free space a recoil momentum $\hbar k_L$ results in a kinetic energy change of $\omega_\mathrm{rec}$ (typically a few kHz).
In a stiff trap $(\omega_\mathrm{rec} \ll \omega_m)$ it is  unlikely that a phonon can be excited due to the insufficient energy associated with the recoil.
In that case, it is well-known \cite{wine,wine2} that the probability of exciting a phonon due to single-photon scattering is suppressed as $\omega_\mathrm{rec}/\omega_m=\eta_\mathrm{LD}^2$. 
\begin{figure}
\centering
\includegraphics[width=\textwidth]{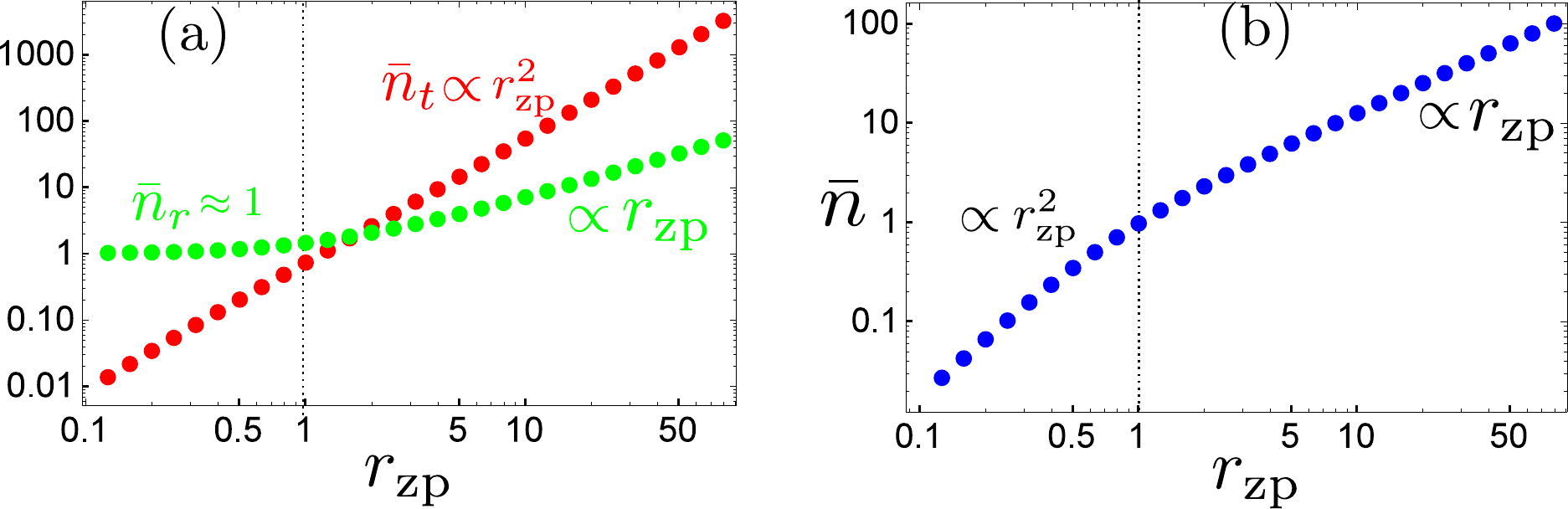}  
\caption{\textbf{Added phonons per photon}\\
We assume critical coupling, the atom to be initially trapped in its ground state and that the resonance position matches with the trap equilibrium ($x_r = x_0$).
\textbf{a)} Conditional  expectation values of created phonons after scattering a single photon $\bar{n}_\mathrm{r/t}$. 
We find the scalings $\bar{n}_r \approx 1$ for $r_\mathrm{zp} \ll 1$ and $\bar{n}_r \propto r_\mathrm{zp}$ for $r_\mathrm{zp} \gg 1$ and $\bar{n}_t \propto r_\mathrm{zp}^2$ for all values of $r_\mathrm{zp}$, leading to a very large number of added phonons for resolved zero-point motion in the case of a measured transmitted photon.
\textbf{b)} Total expectation value $\bar{n}$ (unconditional) of added phonons per photon as a function of  $r_\mathrm{zp}$. The scalings $\bar{n} \propto r_\mathrm{zp}^2$ for $r_\mathrm{zp}\ll 1$ and $\bar{n} \propto r_\mathrm{zp}$ for $r_\mathrm{zp} \gg 1$ originate from the combination of a) and figure~\ref{intuitive}(c), as $\bar{n} = p_t \bar{n}_t + p_r \bar{n}_r$.
}  
\label{totaln}
\end{figure}
However, here we show that for atoms trapped inside cavities,
and in the regime of strong optomechanical coupling, it is possible for a single scattered photon to produce a much larger heating effect, even when the atom is trapped tightly within the Lamb-Dicke limit ($\eta_\mathrm{LD} \ll 1$). The origin of this effect can already be inferred from figure~\ref{entangled}(b), where the post-scattering atomic wave function is seen to be far from the original ground-state wave function due to the narrow spatial features induced by scattering.

In figure~\ref{totaln}(a) we plot the conditional expectation values $\bar{n}_\mathrm{r/t} = \bra{\Psi_\mathrm{r/t}} b^\dag b \ket{\Psi_\mathrm{r/t}}$ of created phonons as a function of $r_\mathrm{zp}$ after measuring a reflected/transmitted photon, respectively. For these plots we assume the atom to be initially in its ground state and that the resonance position matches with the trap equilibrium ($x_r = x_0$).
We find that $\bar{n}_r \approx 1$ for $r_\mathrm{zp} \ll 1$, which reflects the fact that the resulting conditional wave function in this regime is a single-phonon Fock state, as explained in section \ref{ent}. For  $r_\mathrm{zp} \gg 1$ we observe a scaling of $\bar{n}_r \propto r_\mathrm{zp}$, whereas $\bar{n}_t \propto r_\mathrm{zp}^2$ for all values of $r_\mathrm{zp}$. 
We now want to give the intuition behind these scalings. Generally, the number of created phonons is the energy increase normalized with trap frequency: $\bar{n} = \frac{\Delta E}{\omega_m}$.
 The main contribution of added energy comes from the increase in momentum uncertainty, due to the narrow spatial features associated with the conditional wave functions after photon scattering (see figure~\ref{entangled}(b)). Thus, the added energy after one scattering event is approximately 
$
 \Delta E \approx \frac{\bra{\Psi} p^2 \ket{\Psi}}{2m}. 
$
Transmitting a photon localizes the atomic wave function around the resonant position $x_r$ up to an uncertainty of 
  $\Delta x \sim \hbar /\Delta p \sim 1/r_\mathrm{zp}$, which yields a kinetic energy increase corresponding to $\bar{n}_t  \propto r_\mathrm{zp}^2$. The scaling $\bar{n}_r  \propto r_\mathrm{zp}$ for $r_\mathrm{zp} \gg 1$
  is best understood for the case $\kappa_\mathrm{t}=0$ (but the argument holds generally). There, the photon experiences a phase shift $\Phi(x) = \mathrm{arg}[S_r(x)] \approx \arctan[(2(x-x_r)R)/(R^2-(x-x_r)^2)]$ which depends on the atomic position.  $\Phi(x)$ only varies significantly for displacements smaller than $\delta x \lesssim R \propto 1/r_\mathrm{zp}$ and its slope reaches a maximum value of $\Phi'(x_r) \propto r_\mathrm{zp}$.
  The phase shift dominates the contribution to the added kinetic energy, $ \bar{n}_r \propto
 \bra{\Psi} p^2 \ket{\Psi} \propto \int dx |\Psi_0(x)|^2 (\Phi'(x))^2 \propto \int dx (\Phi'(x))^2 \propto  r_\mathrm{zp}^2/r_\mathrm{zp} = r_\mathrm{zp}$ as for $r_\mathrm{zp} \gg 1$, $(\Phi'(x))^2$ peaks over a region much smaller than the width of the wavefunction, and has a width $\propto  1/r_\mathrm{zp}$ and a maximum value of $\propto r_\mathrm{zp}^2$.
 
In figure~\ref{totaln}(b) we plot the unconditional number of added phonons per photon $\bar{n}$ (the photon is not measured after the interaction). 
As it is given by
$
 \bar{n} = p_t \bar{n}_t+ p_r \bar{n}_r
$,
it can be understood as a combination of figure~\ref{totaln}(a) and \ref{intuitive}(c). Thus, the scaling of $\bar{n}_t$ dominates for $r_\mathrm{zp} \ll 1$, whereas the scaling of $\bar{n}_r$ dominates for $r_\mathrm{zp} \gg 1$. 
\subsection{Conclusion}
We have presented the theory of strong optomechanical coupling in nano/micro-cavities, where naturally the mechanical sidebands are unresolved. Possible candidate platforms are trapped atoms or ions in photonic crystal cavities or fiber cavities. We show that these platforms already reach a regime where the atomic zero-point motion is resolved by incident photons, leading to strong entanglement between the photon and the atomic motion. Signatures of this entanglement can be measured in the reflection spectrum, the second-order photon correlation functions, or in the number of added phonons per photon.
Furthermore, we showed that one can create non-Gaussian motional states from Gaussian states by reflecting a single photon, even for unresolved zero-point motion. 
Generally we want to emphasize that the presented theory of this work is relevant to any experiment where atoms are strongly coupled to cavities with small mode volumes.
\ack
 Lukas Neumeier acknowledges financial support from the Spanish Ministry of Economy and Competitiveness, through the ``Severo Ochoa'' Programme for Centres of Excellence in R\&D (SEV-2015-0522). DEC acknowledges support from the Severo Ochoa Programme, Fundacio Privada Cellex, CERCA Programme / Generalitat de Catalunya, ERC Starting Grant FOQAL, MINECO Plan Nacional Grant CANS, MINECO Explora Grant NANOTRAP, and US ONR MURI Grant QOMAND.
 
\clearpage
\appendix

\section{From the Jaynes-Cummings model including motion to an effective model of motion only}
Equation~\ref{fullmodel} of the main text describes the full master equation of a moving two-level atom interacting with a cavity, in the presence of cavity losses and atomic spontaneous emission. In the limit where the cavity is driven near resonantly and the atom is far-detuned, the atomic excited state can be eliminated to yield an effective optomechanical system involving just the atomic motion and the cavity mode. One can go a step further and eliminate the cavity mode, to yield the reduced dynamics of just the atomic motion. The procedure by which a certain degree of freedom can be eliminated from an open system is known as the Nakajima-Zwanzig projection operator formalism \cite{naka,zwanzig,zwerger}, which we now describe here.
\subsection{Projecting out the atomic excited state}\label{ap:atom} 
\begin{figure}
\centering
\includegraphics[width=0.5\textwidth]{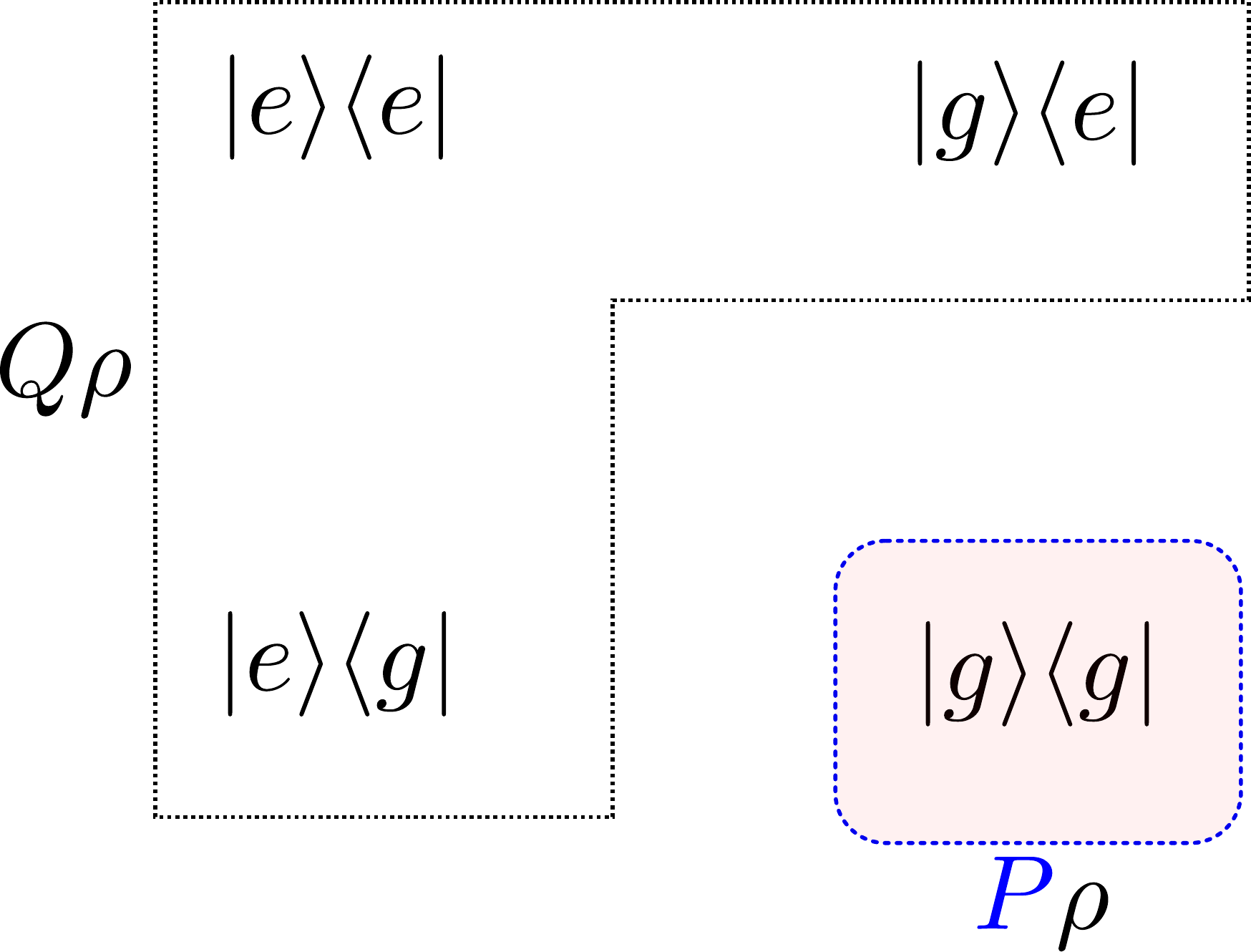} 
\caption{The complete Hilbert space of the internal degrees of freedom of the atom. $P\rho$ is the part we are interested in and the remainder is characterized by the projection operator $Q$.
}
\label{jump0}
\end{figure}

\begin{figure}
\centering
\includegraphics[width=0.7\textwidth]{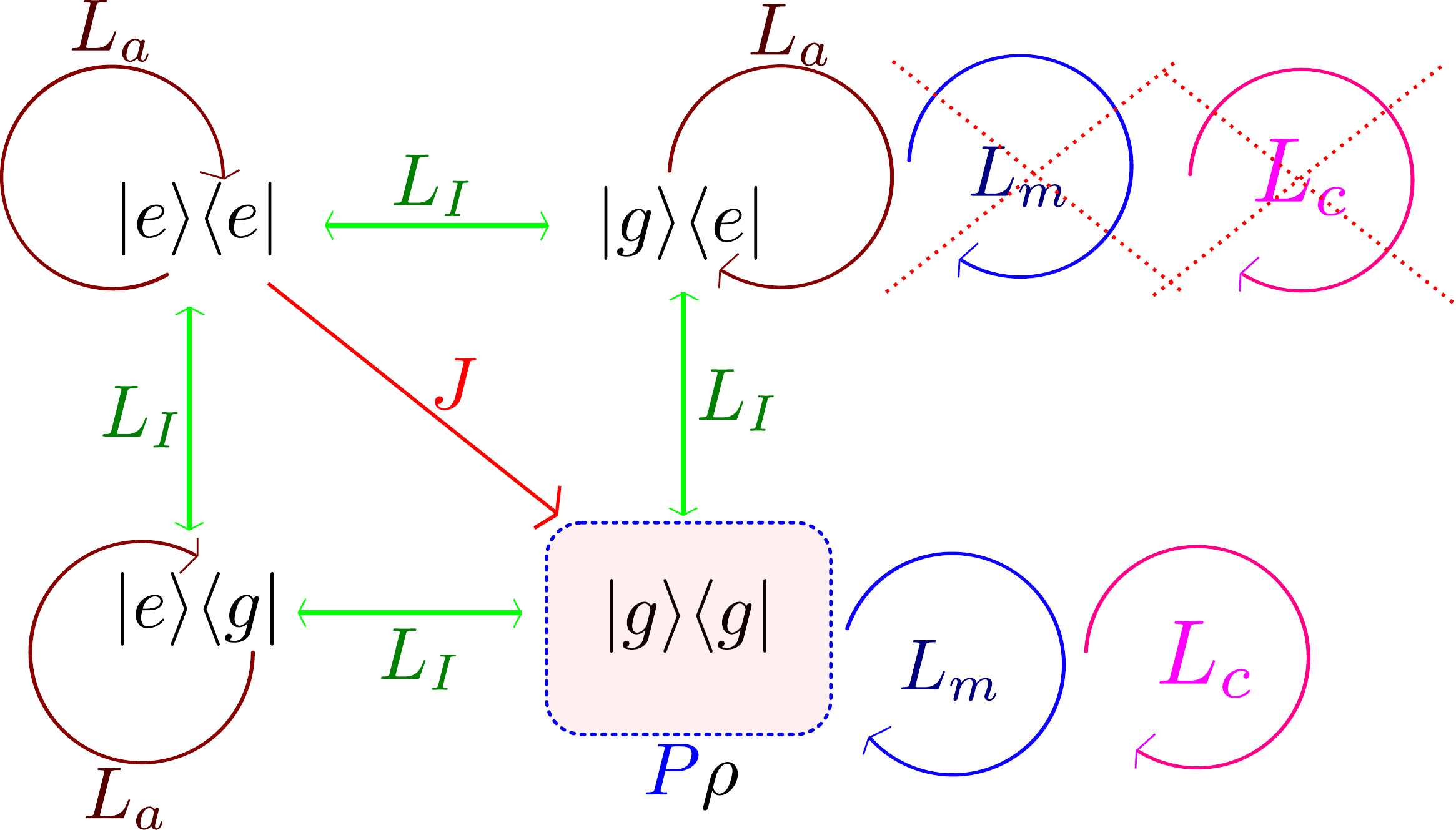} 
\caption{The Hilbert space of the internal degrees of freedom of the atom.
The notation is as follows: The label of an arrow corresponds to a Liouvillian, while the direction of the arrow indicates the possible beginning and ending subspaces of the Liouvillian. For example, the red arrow indicates that the Liouvillian $J$ acting on the subspace $\ket{e}\bra{e}$ takes this subspace to $\ket{g}\bra{g}$.
 Since we assume $\delta_0$ or $\gamma$ to be much larger than $\kappa$ and $\omega_m$, we can neglect the action of $L_o = L_m + L_c$ during a fluctuation out of $P\rho$, which we indicate by crossing them out in the right-top corner and neglecting them in equation~(\ref{ww1}). 
}
\label{jumpfull}
\end{figure}

We first want to eliminate the atomic excited state from the full dynamics of equation~\ref{fullmodel}. It is convenient to define a set of operators $P,Q$, which project the entire system density matrix 
\begin{equation}\label{ro}
 \rho = |g\rangle \langle g| \rho_{gg} +  |g\rangle \langle e| \rho_{ge}+|e\rangle \langle g| \rho_{eg}+|e\rangle \langle e| \rho_{ee},
\end{equation}
into the subspace spanned by $\ket{g}\bra{g}$ (which we want to project the dynamics into), and its orthogonal  $\id-\ket{g}\bra{g}$.
Here $\rho_{ij} =\langle  i |\rho| j \rangle $ are the reduced density matrices for the reduced Hilbert space, which still contain all other existing degrees of freedom. Thus, we define a projection operator $P$:
\begin{equation}
 P \rho = |g\rangle \langle g| \rho_{gg}
\end{equation}
and its complementary
\begin{equation}
 Q \rho =    |g\rangle \langle e| \rho_{ge}+|e\rangle \langle g| \rho_{eg}+|e\rangle \langle e| \rho_{ee}.
\end{equation}
It is straightforward to show $P^2=P,Q^2=Q, QP=0,P+Q = \id$.
In figure~\ref{jump0} we draw a simple picture of the full Hilbert space of the internal degrees of freedom of the atom in order to visualize the part of the Hilbert space we are interested in (described by $P\rho$) and the part we are not (described by $Q\rho)$.  
We will now  divide the super-operator $L$ up in parts 
according to the way they act on the Hilbert space describing the internal degrees of freedom of the atom:
\begin{equation}
 L =L_o + L_a + L_I + J.
\end{equation}
Here, $L_o=L_m+L_c$ is composed of terms that do not act on the internal degrees of freedom, with $L_m$ and $L_c$ describing respectively the trapped atomic motion and the bare dynamics of the driven cavity mode:
\begin{equation}\label{Lm}
 L_m \rho= -\I [\omega_m b^\dag b,\rho]
\end{equation}
\begin{equation}\label{Lc}
L_c\rho =  \I \delta_c [ a^\dag a, \rho ] - \I \sqrt{\kappa_r} E_0  [(a+a^\dag),\rho ] - \frac{\kappa}{2} \left( a^\dag a \rho + \rho a^\dag a - 2 a \rho a^\dag \right).
\end{equation}
The super-operator
\begin{equation}\label{La}
 L_a \rho =  \I  \delta [ \sigma_{ee} ,\rho] - \frac{\gamma}{2} \{ \sigma_{ee}, \rho \}
\end{equation}
acts on $|e\rangle \langle g|,|g\rangle \langle e|,|e\rangle \langle e|$ (the subspace spanned by $Q$) and just multiplies those terms by a c-number. It describes evolution and damping of the excited internal state of the atom.
\begin{equation}\label{Li}
 L_I \rho = - \I [g(x)(\sigma_\mathrm{eg}a + \sigma_\mathrm{ge}a^\dag),\rho]
\end{equation}
acts on all the states and all Hilbert spaces, describing the interaction of the atom with the cavity field and
\begin{equation}\label{J}
 J\rho = \gamma  \sigma_\mathrm{ge} e^{- \I k_c x} \rho e^{\I k_c x} \sigma_\mathrm{eg}
\end{equation}
describes the spontaneous jump of the excited state of the atom into its ground state accompanied by a momentum recoil. In figure~\ref{jumpfull} we draw arrows showing how these super-operators act on different parts of the Hilbert space of atomic internal degrees of freedom.
We are interested in the dynamics of the subspace $P\rho$, while accounting for fluctuations into $Q\rho$. Thus, only  closed loops which start and end in $P\rho$ contribute to the evolution of the reduced density matrix $P\rho$.
To see how this works, we define $v=P\rho$ and $w=Q\rho$ and insert $P+Q = \id$ into equation~(\ref{fullmodel}):
\begin{equation}
 \dot{v} = P\dot{\rho} = P L \rho = P L P \rho+ P L Q \rho. 
\end{equation}
Let us first look at $PLP$:
\begin{equation}
 PLP\rho = P(L_o+L_a + L_I + J)P\rho. 
\end{equation}
To quickly identify vanishing terms we take advantage of figure~\ref{jumpfull} by following the path the super-operators take us through the Hilbert space applying them from the right to the left. Here are some examples:
\begin{enumerate}
 \item The term $PL_IP$: $P$ projects into the subspace $| g \rangle \langle g |$,  while $L_I$ maps a state from $P$ to $Q$. Thus, acting again with $P$ causes this term to vanish.
 \item $PL_aP$: $P$ projects into $| g \rangle \langle g |$ and we immediately see that $L_a$ does not act on it, so this term vanishes.
 \item $PJP = 0$ because $J$ does not act on $| g \rangle \langle g |$. 
 \end{enumerate}
After identifying all vanishing terms, we obtain:
\begin{equation}\label{dd1}
 \dot{v} = L_o v + P (J + L_I) w 
\end{equation}
and
\begin{equation}\label{ww1}
 \dot{w} = QL_Iv + Q (L_o+ L_a + L_I) w. 
\end{equation}
Note that $w$ describes the evolution of the fluctuations out of the subspace of interest. As the timescale of these fluctuations is set by $\delta_0$ and $\gamma$ and we assume that either $\delta_0$ or $\gamma$ is much larger than both $\omega_m$ and $\kappa$, we can neglect the free evolution of the cavity or motion during one of these fluctuations and approximate $L_o w \approx 0$ in equation~(\ref{ww1}), as also indicated in figure~\ref{jumpfull}. Then the general solution to this equation reads:
\begin{equation}
 w(t) = \int_0^t d\tau e^{Q(L_o +L_a)(t-\tau)} Q L_I w(\tau) + \int_0^t d\tau e^{Q(L_o+L_a)(t-\tau)} Q L_I v(\tau)
\end{equation}
where we set $w(0)=0$ as the initial condition.
Now we plug this equation twice into equation~(\ref{dd1}) (iteratively) in order to catch a term of the order $J L_I^2$:
\begin{align}
 \dot{v}(t) & = \nonumber L_o v + P(J +L_I) \int_0^t d\tau e^{Q(L_o +L_a)(t-\tau)} Q L_I v(\tau) \\ & + P(J +L_I)  \int_0^t d\tau e^{Q(L_o +L_a)(t-\tau)} Q L_I \int_0^\tau d\tau' e^{Q(L_o +L_a)(t-\tau')} Q L_I v(\tau').
\end{align}
Here we neglected the term proportional to $w(\tau')$ since it produces only terms $\propto L_I^3$ or higher.
Again by following the path of how these super-operators act with figure~\ref{jumpfull}, we can quickly identify which terms vanish since all contributing terms need to have closed loops starting and ending in $| g \rangle \langle g |$. So we are left with:
\begin{align}\label{las1t}
 \dot{v}(t) & \nonumber = L_o v+ PL_I \int_0^t d\tau e^{(L_o +L_a)(t-\tau)}  L_I v(\tau) \\ & + PJ \int_0^t d\tau e^{(L_o +L_a)(t-\tau)}  L_I \int_0^\tau d\tau' e^{(L_o +L_a)(t-\tau')}  L_I v(\tau').
\end{align}
After extending the lower integral borders to $-\infty$ (Markov approximation), we obtain equation~(\ref{mastaop}) of the main text.
\subsection{Projecting out the cavity field} \label{ap:cav}
\begin{figure}
\centering
\includegraphics[width=0.5\textwidth]{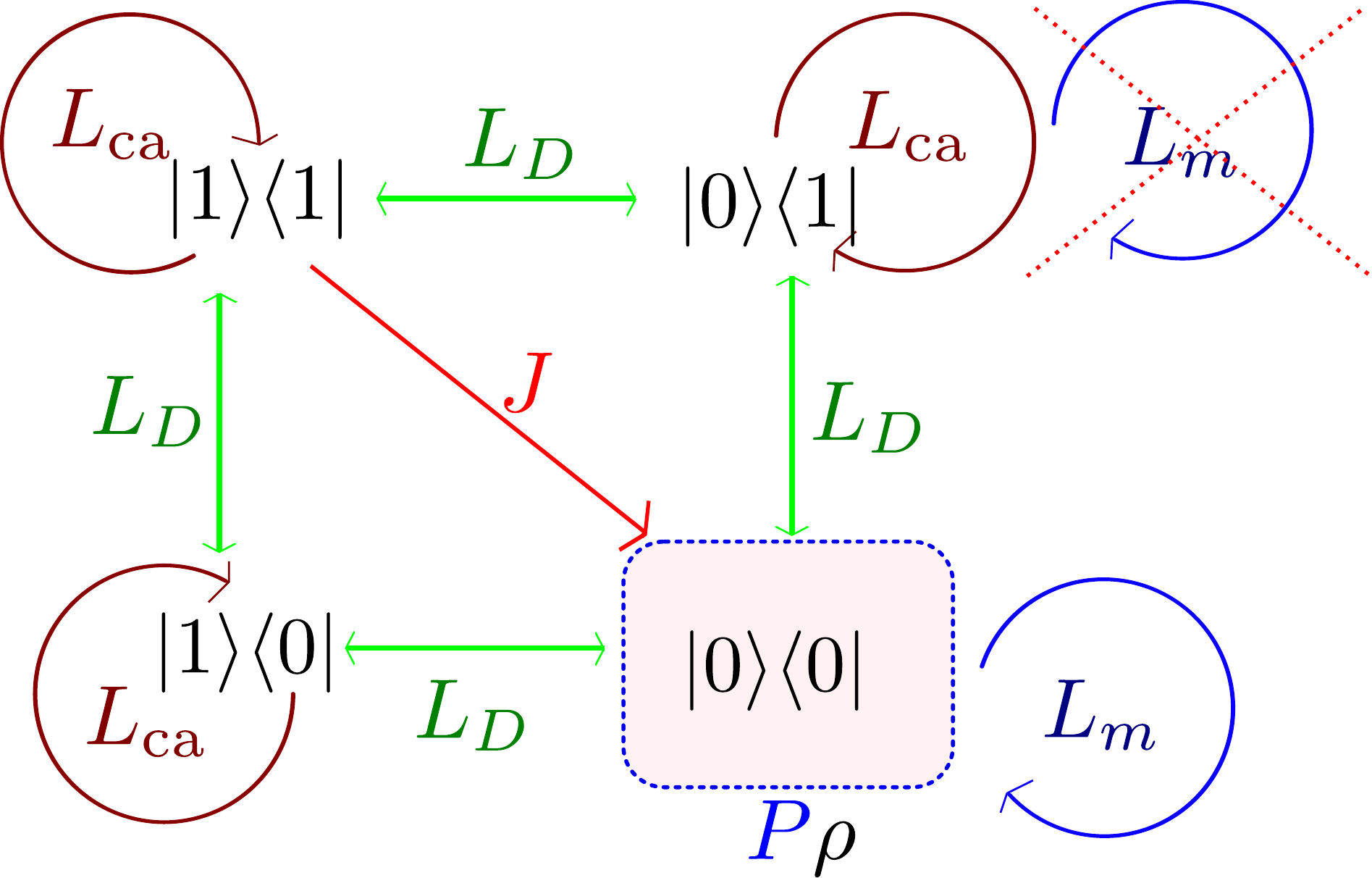} 
\caption{The Hilbert space of the single excitation subspace of the cavity.
The label of an arrow corresponds to a Liouvillian, while the direction of the arrow indicates the possible beginning and ending subspaces of the Liouvillian. For example, the red arrow indicates that the Liouvillian $J$ acting on the subspace $\ket{1}\bra{1}$ takes this subspace to $\ket{0}\bra{0}$.
As we assume $\kappa \gg \omega_m$, we can neglect the time evolution due to the super-operator $L_m$ during a fluctuation out of $P\rho$.
}
\label{cavityraus}
\end{figure}

The next step is to find a master equation only containing motional degrees of freedom ($p$ and $x$) of the atom as operators. In order to find this equation we need to use the Nakajima-Zwanzig technique to project out the cavity mode from  equation~(\ref{mastaop}). For the sake of simplicity we assume $\delta_0 \gg \gamma$ (and thus $ \frac{g_0^2}{ \delta_0^2 + \frac{\gamma^2}{4}} \approx \frac{g_0^2}{\delta_0^2}$) and $\kappa \gg \gamma$ in the following, so we can ignore the atomic decay channel for this derivation by approximating $L_\mathrm{om} \approx L_\kappa$.  
For weak driving, we can restrict ourselves to the photon subspace defined by $|0\rangle$ , $|1\rangle$.
Subsequently, we can adopt our projection operator formalism from above and write the density operator as follows:
\begin{equation}\label{ro}
 \rho = |0\rangle \langle 0| \rho_{00} +  |0\rangle \langle 1| \rho_{01}+|1\rangle \langle 0| \rho_{10}+|1\rangle \langle 1| \rho_{11}
\end{equation}
with $\rho_{ij} =\langle  i |\rho| j \rangle $ being the reduced density matrix describing atomic motion.
As we are interested in the subspace spanned by $|0\rangle \langle 0|$ we define an projection operator $P$:
\begin{equation}
 P \rho = |0\rangle \langle 0| \rho_{00}
\end{equation}
and 
\begin{equation}
 Q \rho =    |0\rangle \langle 1| \rho_{01}+|1\rangle \langle 0| \rho_{10}+|1\rangle \langle 1| \rho_{11}.
\end{equation}
We again decompose the total Liouvillian in parts according to the way they act:  
\begin{equation}
 L = L_m + L_\mathrm{ca} + L_D + J 
\end{equation}
with $L_m$ defined in equation~(\ref{Lm}),
\begin{equation}
 L_\mathrm{ca} \approx - \I [-\Delta(x)a^\dag a, \rho] - \frac{\kappa}{2}\{ a^\dag a, \rho \}
\end{equation}
and $L_D\rho = -\I \sqrt{\kappa_r} E_0  [a + a^\dag, \rho]$,
which describes the interaction of the cavity mode with an external coherent laser drive.
$
 J\rho = \kappa a \rho a^\dag 
$
describes the spontaneous decay of the cavity mode.
Now we draw in figure~\ref{cavityraus} a picture of the Hilbert space of the degrees of freedom of the cavity, including the arrows which illustrate how these defined super-operators act.
A similar prodecure as in \ref{ap:atom} leads to the quantum master equation (\ref{nmastacon}) of the main text describing atomic motion.
\section{Single Photon Scattering Theory}\label{ap:scat}
Here we provide details of the derivation of equations (\ref{smex}) and (\ref{sm0}) in the main text. Inserting equations (\ref{in}) and (\ref{out}) into equation~(\ref{ss}) and multiplying with $\bra{(\omega')_\mathrm{r/t},m}$ from the left gives us an equation for the S-matrix elements:
\begin{equation}
 S_\mathrm{r/t,n}(\omega_L) \delta(\omega_L-\omega'- n \omega_m) = \bra{(\omega')_\mathrm{r/t},n }S \ket{(\omega_L)_\mathrm{left},0}
\end{equation}
where $\omega'$ refers to the frequency of the reflected or transmitted photon.
In the following, we will establish a connection between the S-matrix elements, and the standard input-output formalism of cavity QED \cite{marco}. Conveniently, this connection enables one to calculate S-matrix elements based upon knowledge of the eigenvalues and eigenstates of the system Hamiltonian $H_\mathrm{eff}$. The input-output equation states that the output field in each decay channel (reflection/transmission) is the sum of the input field and the field emitted by the scattering center. For example the input-output equation for photon reflection is given by 
\begin{equation}\label{inout}
  a_\mathrm{out}(t) = a_\mathrm{in}(t) - \I \sqrt{\kappa_r} a(t)
\end{equation}
where for notational convenience we leave out the subscript ``r'' in the input and output ports. The scattering operators $a_\mathrm{in/out}(\omega)$ are connected to the input-output Heisenberg-Langevin operators $a_\mathrm{in/out}(t)$  by a simple Fourier transform \cite{shan}
\begin{equation}
 a_\mathrm{in/out}(\omega) = \frac{1}{\sqrt{2 \pi}} \int dt e^{\I \omega t} a_\mathrm{in/out}(t).
\end{equation}
 Now we focus on the S-matrix for the process of photon reflection
\begin{equation}
 S_\mathrm{r,n}(\omega_L) \delta(\omega_L-\omega'- n \omega_m)= \bra{0_c,n} a_\mathrm{out}(\omega') a_\mathrm{in}^\dag (\omega_L) \ket{0_c,0}
\end{equation}
where we expressed the S-matrix in terms of scattering operators $a_\mathrm{in}^\dag (\omega_L)$ and $a_\mathrm{out}(\omega')$ which create in- and out-going monochromatic scattering states \cite{scatteringshit}.
Using the input-output equation, one can re-write $a_\mathrm{out}$ in terms of the cavity field and input field, yielding
\begin{equation}
 S_\mathrm{r,n}(\omega_L)\delta(\omega_L-\omega'- n \omega_m)= \delta(\omega_L-\omega') \delta_{n,0} - \I \sqrt{\kappa_r} \bra{0_c,n} a(\omega') a_\mathrm{in}^\dag (\omega_L) \ket{0_c,0}. 
\end{equation}
Now we replace the scattering operators with the Fourier transform of the corresponding input-output operators. The matrix element $\bra{0_c,n} a(t') a_\mathrm{in}^\dag (t_L) \ket{0_c,0}$ vanishes for $t_L > t'$ since $[a(t'),a^\dag_\mathrm{in}(t_L)] = 0$ for $t_L > t'$ and $\bra{0_c} a^\dag_\mathrm{in}(t_L) = 0$. Thus, we introduce the time ordering operator $T$ making sure that $t' > t_L$. Then we have
\begin{equation}
\bra{0_c,n} T[a(t') a_\mathrm{in}^\dag (t_L)] \ket{0_c,0} = 
- \I \sqrt{\kappa_r}  \bra{0_c,n} T[a(t')  a^\dag(t_L)]  \ket{0_c,0}, 
\end{equation}
where we replaced $a_\mathrm{in}(t_L)$ with $a(t_L)$ using the input-output equation. The term containing the output operator vanishes as $[a(t'),a^\dag_\mathrm{out}(t_L)]=0$ for $ t' > t_L$ (which is already ensured by T) and $\bra{0_c} a^\dag_\mathrm{out}(t_L) = 0$.
Finally, we arrive at 
\begin{equation}\label{srn}
 S_\mathrm{r,n}(\omega_L)\delta(\omega_L-\omega'- n \omega_m)= \delta(\omega_L-\omega') \delta_{n,0} - \kappa_r \tau_\mathrm{n}(\omega_L)
\end{equation}
with 
\begin{equation}\label{tau}
 \tau_\mathrm{n}(\omega_L) = \frac{1}{2 \pi} \int dt_L dt' e^{\I(\omega' t' - \omega_L t_L)} \bra{0_c,n} T a(t') a^\dag(t_L) \ket{0_c,0}.
\end{equation} 
For the S-matrix describing the process of photon transmission we obtain
\begin{equation}\label{srt}
 S_\mathrm{t,n}(\omega_L) \delta(\omega_L-\omega'- n \omega_m)=  - \sqrt{\kappa_r\kappa_t }\tau_\mathrm{n}(\omega_L)
\end{equation}
Note that the S-matrix of reflection $S_r$  includes the term $\delta(\omega_L-\omega') \delta_{n,0}$ describing interaction-free reflection of photons. In contrast, in the S-matrix of transmission $S_t$ there is no such term, since the input field on the transmitting side of the cavity is in the vacuum state and thus the transmitted field is built exclusively from the emission of photons by the scattering center.
We can write
\begin{equation}
 \langle a(t') a^\dag(t_L) \rangle = \mathrm{Tr} \left[ a e^{L_s(t'-t_L)}a^\dag \rho(0) \right],
\end{equation}
where $\rho(0) = \ket{0_c,0} \bra{0_c,0} $ and
$L_s \rho = - \I [H_\mathrm{eff},\rho] + \kappa a \rho a^\dag$ with $H_\mathrm{eff}$ described by equation~(\ref{int}) from the main text. Since the term $\kappa a \rho a^\dag$ reduces the number of photons, its contribution vanishes as the correlator conserves the number of photons.
Thus, the evolution of $a(t)$ is governed by $H_\mathrm{eff}$ alone and for evaluating the S-matrix we can effectively use
\begin{equation}
 a(t) = e^{\I H_\mathrm{eff} t} a e^{-\I H_\mathrm{eff} t}. 
\end{equation} 
We further express
\begin{equation}
  \bra{0_c,n} T a(t') a^\dag(t_L) \ket{0_c,0} = \Theta(t_L-t') e^{\I \omega_n n t_L} \bra{1_c,n} e^{-\I H_\mathrm{eff}(t_L-t')} \ket{1_c,0}
\end{equation}
where $e^{\I \omega_n n t_L}$ counts the energy of the created phonons during the scattering process  and the step function $\Theta(t_L-t')$ which vanishes for $t_L<t'$ ensures time ordering. In order to express the S-matrix fully in terms of eigenvalues $\lambda_\beta$  and eigenstates $\ket{\beta}$ of $H_\mathrm{eff}$ with  $H_\mathrm{eff} \ket{\beta} = \lambda_\beta \ket{\beta}$  we insert a unity operator $1= \sum_\beta \ket{\beta}\bra{\beta}$
right before $\ket{1_c,0}$.
Therefore we write 
\begin{equation}\label{unitss}
 \bra{1_c,n} e^{-\I H_\mathrm{eff}(t_L-t')} \ket{1_c,0} = \sum_\beta    \langle 1_c,n|\beta \rangle e^{-\I \lambda_\beta (t_L-t')}  \langle \beta|1_c,0 \rangle  
\end{equation}
where $\langle 1_c,n|\beta \rangle $ is the projection of the eigenstates $\ket{\beta}$ into the basis states $\bra{1_c,n}$.
After evaluating the Fourier transform in equation~(\ref{tau}) we are left with
\begin{equation}
 \tau_\mathrm{n}(\omega_L) = - \I \delta(\omega_L - \omega' - n \omega_m) \sum_{\beta}  \langle 1_c,n|\beta \rangle \frac{1}{\lambda_\beta}  \langle \beta|1_c,0 \rangle.
\end{equation}
which together with equation~(\ref{srn}) and (\ref{srt}) reproduces equation~(\ref{smex}) and (\ref{sm0}) in the main text. 
\section{The full effective theory and its validity} \label{ap:full}
Here we begin by generalizing our effective theory presented in the main text (Section \ref{masterjunge} and \ref{scattar}) by including spontaneous emission into the master equation (\ref{nmastacon}) and the single photon scattering output state (\ref{state}). Then we define the parameter space for which our theory is valid.
We do this by comparing results of our effective theory with a numerical simulation of the full Jaynes-Cummings model including motion (\ref{fullmodel}) where the only assumption is the Lamb-Dicke regime $\eta_\mathrm{LD} \ll 1$ which allows for the linearization of the mode profile $u(x)$. This approximation is only done for numerical purposes and we note that our effective theory does not depend on the Lamb-Dicke parameter.

For systems where $\kappa \gg \gamma$ is not true, we need to include the atomic decay channel.
Doing so, the single photon scattering output state now generalizes to:
\begin{equation}\label{outfull}
 \ket{\Psi_\mathrm{out}} = S_r(\omega_L,x) \Psi_\mathrm{0}(x)\ket{(\omega_L)_r} + S_\mathrm{t}(\omega_L,x) \Psi_0(x)  \ket{(\omega_L)_\mathrm{t}} + S_\mathrm{at}(\omega_L,x) \Psi_0(x)\ket{(\omega_L)_\mathrm{at}}
\end{equation}
where the scattering matrices for reflection, transmission and the scattering matrix for spontaneous emission are respectively given by:
\begin{equation}\label{sexf}
  S_r(\omega_L,x) =  1 -   \frac{\I \kappa_r}{\Delta_c(x) + \I \frac{\kappa(x)}{2}}
\end{equation}
\begin{equation}\label{s0f}
  S_{\mathrm{t}}(\omega_L,x) =  -   \frac{\I\sqrt{\kappa_\mathrm{t} \kappa_r} }{\Delta_c(x) + \I \frac{\kappa(x)}{2}},
\end{equation}
\begin{equation}\label{at}
 S_\mathrm{at}(\omega_L,x) = \sqrt{\frac{g_0^2}{\delta_0^2 + \frac{\gamma^2}{4}}} \frac{\I \sqrt{\gamma \kappa_r}}{\Delta_c(x) + \I \frac{\kappa(x)}{2}} u(x)  e^{\I k_c x}.
\end{equation}
The scattering matrices conserve probability and obey $|S_r(\omega_L,x)|^2 + |S_t(\omega_L,x)|^2  + |S_\mathrm{at}(\omega_L,x)|^2 = 1$ for all values of $\omega_L$ and $x$ . 
Note that we treat here for simplicity only one direction of spontaneous emission which has a one dimensional decay channel described by $\ket{(\omega_L)_\mathrm{at}}$. The resulting momentum kick qualitatively reproduces the main effect that would occur in a full three-dimensional treatment of spontaneous emission. We also did not exclusively account for intrinsic cavity losses at a possible rate $\kappa_\mathrm{in}$, however including this process  would simply result in an additional term in the output state equation~(\ref{outfull}) with a corresponding S-matrix that looks like $S_t$, but with $\kappa_t$ replaced by $\kappa_\mathrm{in}$.
The total effective linewidth of the cavity is increased by the effective rate of spontaneous emission 
\begin{equation}\label{ka2}
 \kappa(x) = \kappa_r + \kappa_t + \gamma  \frac{g_0^2}{\delta_0^2+ \frac{\gamma^2}{4}} u^2(x), 
\end{equation}
which depends on the position of the atom. 
As explained in the main text, we can express the jump operators in terms of the scattering matrices such that they describe intuitive physical decay processes. The corresponding master equation describing a coherent drive is then given by:
\begin{equation}\label{mag}
 \dot{\rho} = -\I (H_e\rho- \rho H_e^\dagger) + E_0^2 (S_r\rho S_r^\dag + S_t\rho S_t^\dag + S_\mathrm{at} \rho S_\mathrm{at}^\dag) 
\end{equation}
with the Hamiltonian
\begin{equation}\label{he}
 H_e = \omega_m b^\dag b - \frac{\I}{2} E_0^2.
\end{equation}
Note that by including spontaneous emission into the model the zero-point resolution reads in good approximation
 \begin{equation}\label{rzp}
  r_\mathrm{zp} \approx \eta_\mathrm{LD} \frac{2 g_0^2 |\delta_0|}{\kappa(x_0)  (\delta_0^2 + \frac{\gamma^2}{4})}.
\end{equation}
We have averaged the position dependent effective decay rate  $\kappa(x_0) \approx \bra{\Psi_0} \kappa(x) \ket{\Psi_0}$ with the atomic wave function $\Psi_0(x)$.

In order to derive the single photon output state (\ref{outfull}) and the master equation (\ref{mag}) we made two assumptions:
\begin{enumerate}
 \item Large atom/laser detuning $\delta_0 \gg g_0$, which allowed us to effectively eliminate the excited state of the atom leading to an effective optomechanical master equation (\ref{mastaop}). Note that a large spontaneous emission rate $\gamma \gg g_0$ would allow this elimination as well. However, here we are interested in strongly coupled systems, where $g_0 \gtrsim \gamma$.
 \item Unresolved vibrational sidebands $\kappa \gg \omega_m$ which allowed us to derive the output state (\ref{outfull}) and effectively eliminate the cavity mode in order to derive the master equation (\ref{mag}).
\end{enumerate}
Now we will check the limits of these assumptions by numerically simulating a single photon scattering event with the full model (equation~(\ref{fullmodel})).
The numerical simulation is done by diagonalizing the Hamiltonian
\begin{align}\label{num}
 H_\mathrm{D} = \omega_m b^\dag b - (\delta_0 + \I \frac{\gamma}{2}) \sigma_\mathrm{ee} -(\delta_c + \I \frac{\kappa}{2})a^\dag a  + g_0(u(x_0)+ g_0 \eta_\mathrm{LD} (b^\dag + b))(a^\dag \sigma_\mathrm{ge} + h.c.),
\end{align}
 in the single-photon subspace and using the eigenvalues and eigenstates in the exact scattering matrices for reflection, transmittion and atomic decay constructed according to equation~(\ref{smex}) and equation~(\ref{sm0}) which is described in \ref{ap:scat}. One has to take care that the unity operator as inserted in equation~(\ref{unitss}), is here $\id = \sum_\beta \ket{\beta}\bra{\beta^*}$, with the eigenvectors normalized as $\braket{\beta^*}{\beta}=1$, since the Hamiltonian $H_D$ is complex symmetric due to losses rather than Hermitian.
\subsection{Limits of the assumption $|\delta_0| \gg g_0$ }
We begin with the question of how large $\frac{g_0}{|\delta_0|}$ can be, such that all approximations previously made are still valid. This is important to know, as the previously studied regime of resolved zero point motion $r_\mathrm{zp} \gg 1$ requires a large effective optomechanical coupling $r_\mathrm{zp} \propto g_{\mathrm{om}} \appropto \frac{g_0}{|\delta_0|}$. Thus, to reach this regime, it is beneficial to choose  $\frac{g_0}{|\delta_0|}$ as large as possible. However, increasing this fraction, we will eventually leave the parameter space in which our effective theory correctly predicts results.  
\begin{figure}
\centering
\includegraphics[width=\textwidth]{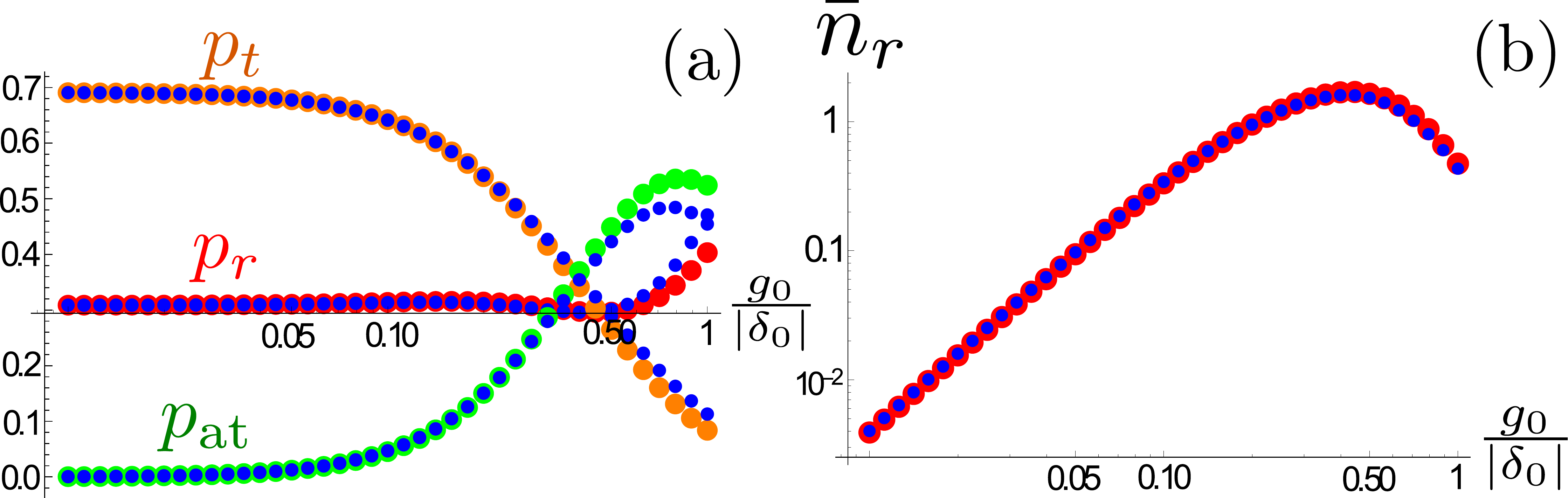} 
\caption{ \textbf{Effective theory vs numerical simulation.}
We assume the atom to be initially in its motional ground state and that the incident photon is on resonance with the atom-cavity system.\\
\textbf{a)} Probability of photon reflection $p_r$ (red), photon transmission $p_t$ (orange) and spontaneous emission $p_\mathrm{at}$ (green) as a function of $\frac{g_0}{|\delta_0|}$ and calculated with the effective theory. Blue smaller dots: exact numerical simulation. Here, we have used parameters from a recente fiber cavity experiment with trapped $^{40}\mathrm{Ca}^+$-ions (see \ref{ap:cryst}.2, parameter set II). 
Here we choose $\kappa_t = 2 \pi \times 0.8 \,\mathrm{MHz}$, $\kappa_r = 2 \pi \times 2.8 \, \mathrm{MHz}$, $\eta_\mathrm{LD}= \sqrt{\omega_\mathrm{rec}/\omega_m} = 0.2$, $\omega_m = 2 \pi \times 0.2 \, \mathrm{Mhz}$.\\
\textbf{b)} Conditional phonon expectation value $\bar{n}_r$ given that a photon is reflected from the cavity for the same parameters as a). The effective theory (red) matches very well with the numerical simulation (blue).
}
\label{approxps}
\end{figure}
To understand when this happens we will now compare our effective theory with a numerical simulation of the full master equation (\ref{num}) as a function of $g_0/|\delta_0|$ (and later as a function of $\omega_m/\kappa$ for similar reasons).
We will assume in the following that the atom is trapped in its motional ground state at a location with maximum intra cavity intensity slope $k_c x_0 = \pi/4$ and, if not stated otherwise, that the single incident photon is on resonance with the atom-cavity system $\Delta_c(x_0) = 0$, which implies $x_r = x_0$.
Figure~\ref{approxps}(a) shows the probability of photon reflection $p_r$ (red), photon transmission (orange) and spontaneous emission $p_\mathrm{at}$ (green) as a function of $g_0/|\delta_0|$ calculated with the effective theory:
\begin{equation}\label{refl1}
 p_\mathrm{r/t/at}(\omega_L) = \int dx |S_\mathrm{r/t/at}(\omega_L,x)|^2 |\Psi_0(x)|^2.
\end{equation}
We use for $|S_\mathrm{r/t/at}(\omega_L,x)|^2$, equation~(\ref{sexf}), equation~(\ref{s0f}) and equation~(\ref{at}), respectively. We also use parameters from a recent fiber cavity experiment (\ref{ap:cryst}.2), where $\gamma > \kappa$ and thus, one needs to account for spontaneous emission.
The blue dots correspond to the full numerical simulation of the Jaynes-Cummings model including motion (equation~(\ref{num})). We observe a great match for $g_0/|\delta_0| < 1/2$. Figure~\ref{approxps}(b) shows the conditional phonon expectation value  $\bar{n}_r = \bra{\Psi_r} b^\dag b \ket{\Psi_r}$ given a reflected photon as a function of $g_0/|\delta_0|$ for the same parameters as a). $\Psi_r(x)$ is given by equation~(\ref{psir}) in the main text. We observe a great match for $g_0/|\delta_0| < 1$. 

\subsection{Limits of the assumption $\kappa \gg \omega_m$}
\begin{figure}
\centering
\includegraphics[width=0.6\textwidth]{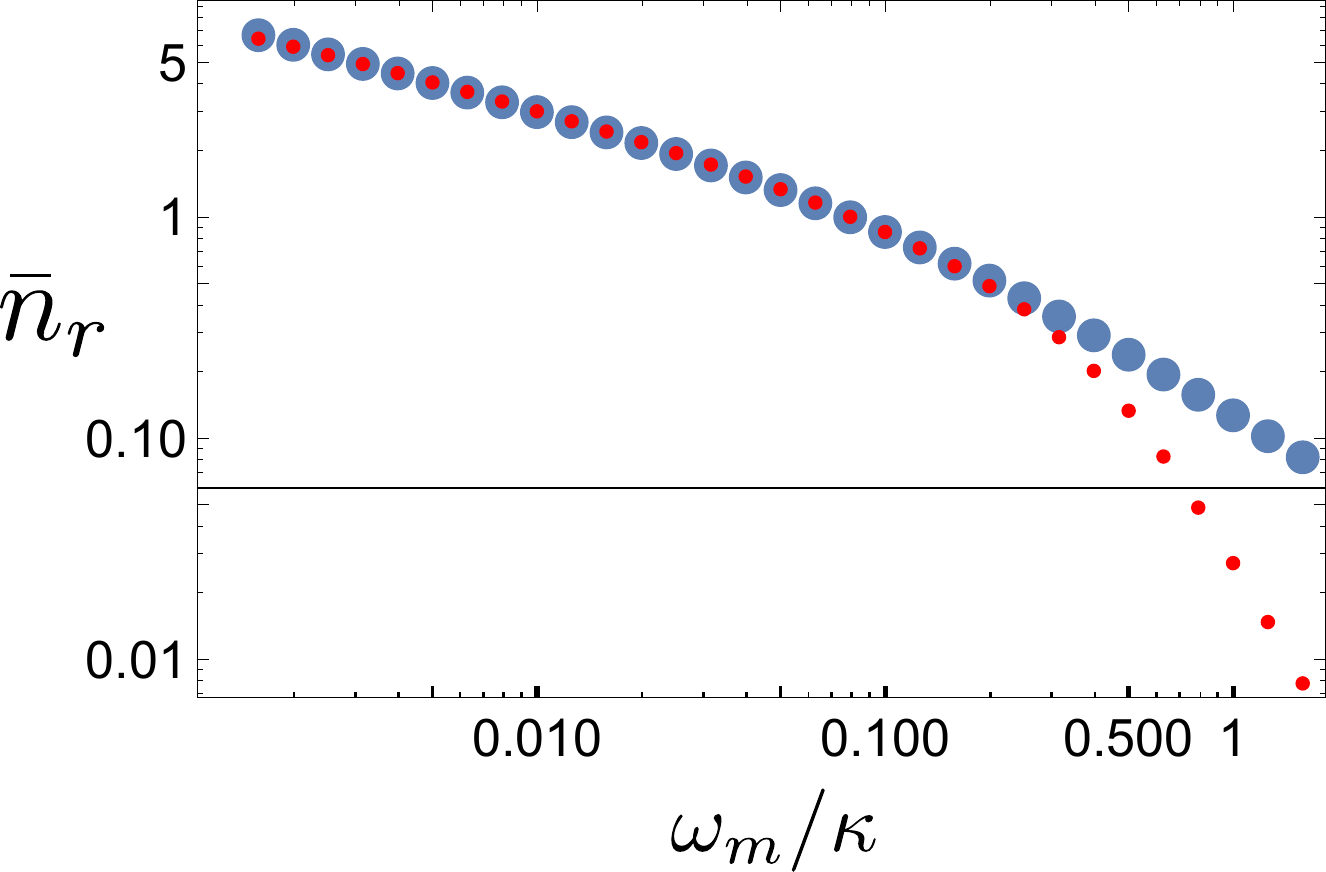} 
\caption{ \textbf{Effective theory (blue) vs numerical simulation (red dots) approaching sideband resolution}  \\
We assume the atom to be initially in its motional ground state and that the incident photon is on resonance with the atom-cavity system.
We plot the phonon expectation value $\bar{n}_r$ after reflecting a photon as a function of $\omega_m/\kappa$.  Parameters are chosen for an atom trapped inside a photonic crystal cavity as presented in \ref{ap:cryst}.1. We choose an atom-cavity detuning of $\omega_0-\omega_c =100 g_0$  and an artificial value of $\kappa=2 \pi \times 20 \, \mathrm{MHz}$ (with $\kappa_r=4 \kappa_t$) as we only want to check the validity of the effective theory once sideband resolution is approached.}
\label{approx}
\end{figure}

Here we want to check the validity of the effective theory once sideband resolution is approached.
We plot the created phonon expectation value $\bar{n}_r$ after reflecting a single photon in figure~\ref{approx} as a function of $\frac{\omega_m}{\kappa}$.  Here, we take the vacuum Rabi splitting $g_0=2 \pi \times 10 \, \mathrm{GHz}$ corresponding to a possible photonic crystal cavity (\ref{ap:cryst}.1), an atom-cavity detuning of $\omega_0-\omega_c=100g_0$, and again we consider a resonant photon for an atom trapped at $k_c x_0 = \pi/4$. For illustrative purposes, we take an artificially low value of $\kappa=2 \pi \times 20 \, \mathrm{MHz}$, which is distributed only between reflection and transmission ports (with $\kappa_r=4 \kappa_t$), and allow $\omega_m$ to vary. We observe a reasonable match between the exact numerical simulation and our effective model for $\omega/\kappa < 1/4$.
\section{Experimental canditate systems for resolving zero-point motion} \label{ap:cryst}
\subsection{Photonic Crystal Cavities} 
The coupling of atoms to the mode of a photonic crystal cavity can be as large as $g_0 \sim 2 \pi \times 10 \, \mathrm{GHz}$ \cite{couplingg0} for Rubidium atoms. Rubidium atoms have a natural linewidth of $\gamma \sim 2 \pi \times 6 \, \mathrm{MHz}$ and a recoil frequency of $\omega_\mathrm{rec} \approx 2 \pi \times 3.8 \, \mathrm{kHz}$ for a resonant photon wavelength around $\lambda_c \approx 780 \mathrm{n m}$. At the same time quality factors of more than $Q \sim 10^6$ are feasible inside photonic crystal nano-cavities \cite{crystal}, associated with a decay rate of rougly $\kappa \sim 2 \pi \times 0.25  \, \mathrm{GHz}$. Since $\gamma \ll \kappa$, 
spontaneous emission can be ignored and experiments are very well described by the effective master equation (equation~(\ref{nmastacon})) and the effective output state (equation~(\ref{state})). The achievable zero-point resolution in photonic crystal cavities is $r_\mathrm{zp} \sim 10$ by taking $\eta_\mathrm{LD}=0.25$ (calculated with equation~(\ref{rzp})).
\subsection{Fiber Cavities} \label{ap:fiber}
Here we discuss a fiber cavity QED experiment with trapped $^{40}\mathrm{Ca}^+$-ions ($\omega_\mathrm{rec}\approx 2 \pi \times 6.8 \, \mathrm{kHz}$, $\gamma = 2 \pi \times 11.2 \, \mathrm{MHz}$) by Tracy Northup in Innbruck \cite{tracy0}.
They are able to realize different sets of $g_0$ and $\kappa$ by changing the cavity length. Here we give two examples:
\begin{enumerate}
 \item Parameter set I is given by: $g_0 = 2 \pi \times 41 \, \mathrm{MHz}$ , $\kappa = 2 \pi \times 8 \, \mathrm{MHz}$.
 \item Parameter set II is given by: $g_0 = 2 \pi \times 21 \, \mathrm{MHz}$ and $\kappa = 2 \pi \times 3.6 \, \mathrm{MHz}$.
\end{enumerate}
\begin{figure}
\centering 
\includegraphics[width=\textwidth]{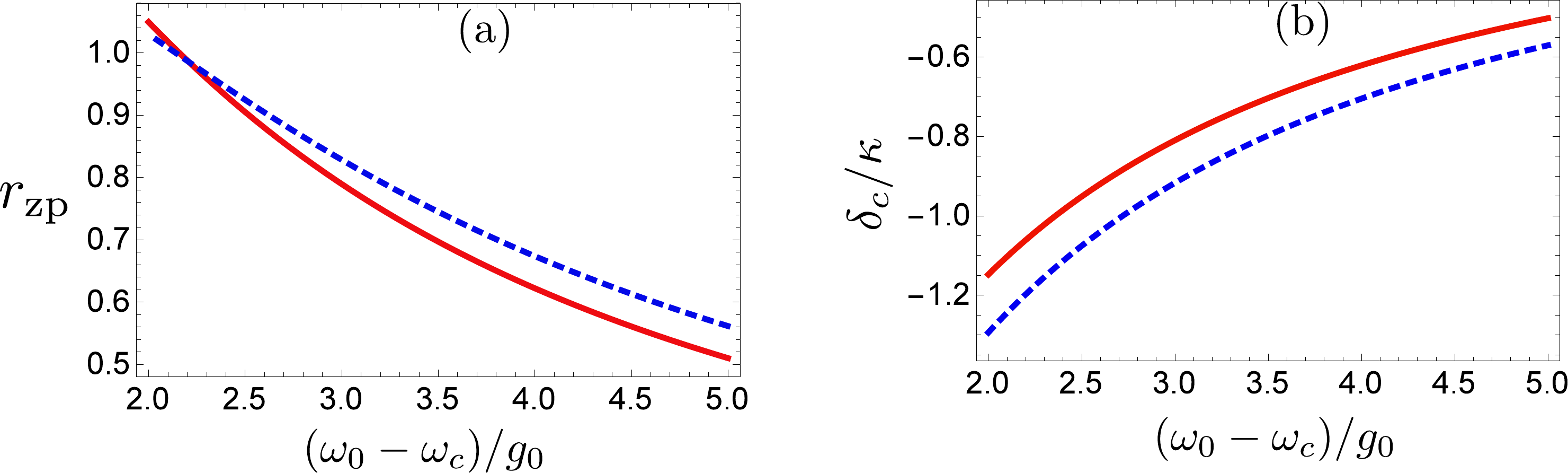} 
\caption{\textbf{a)} Zero-point resolution $r_\mathrm{zp}$ as a function of cavity-atom detuning $\omega_0-\omega_c$ for parameter set I (red) and set II (blue, dashed) of a tunable fiber cavity experiment with trapped ions. For parameters see \ref{ap:fiber}. Here, we choose $\omega_m = 2 \pi \times 0.1 \,\mathrm{MHz}$, $k_c x_0 = \pi/4$ and $\delta_c$ such that $x_r = x_0$ (see b)).
\\
\textbf{b)} Here we show how to choose $\delta_c$ in order to ensure $k_c x_r=k_c x_0= \pi/4$. Plotted is the cavity-laser detuning $\delta_c$ as a function of $\omega_0 - \omega_c$ for parameter set I (red) and set II (blue, dashed) satisfying  the condition $\Delta_c(x_r)=0$.
}
\label{trac11}
\end{figure}
 Figure~\ref{trac11}(a) shows the zero-point resolution $r_\mathrm{zp}$ as a function of cavity-atom detuning $\omega_0-\omega_c$ for parameter set I (red) and set II (blue, dashed) calculated with equation~(\ref{rzp}). We choose $\omega_m = 2 \pi \times 0.1 \,\mathrm{MHz}$, $k_c x_0 = \pi/4$. $\delta_c$ is chosen in a way that the condition $\Delta_c(x_r) = 0$ is satisfied, which implies $x_r = x_0$.
We observe that by choosing $\omega_0-\omega_c = 2 g_0$  one achieves  $r_\mathrm{zp} \approx 1.05$ with parameters set I and $r_\mathrm{zp} \approx 1.03$ with parameter set II. 
We also demonstrate how to choose $\delta_c$ in order to obtain $k_c x_r =k_c x_0 = \pi/4 $ in figure~\ref{trac11}(b), which shows $\delta_c$ as a function of $\omega_0 - \omega_c$ for parameter set I (red) and set II (blue, dashed).

Note that because the spontaneous emission rate $\gamma$ is comparable to $\kappa$, the process of spontaneous emission cannot be neglected and the master equation (\ref{mag}) and single photon scattering output state (\ref{outfull}) need to be applied in order to predict outcomes of this experiment.

\end{document}